\documentclass[journal]{IEEEtran}

\ifCLASSINFOpdf

\else

\fi
\usepackage{graphicx}
\usepackage{amsmath, amssymb, amsthm}
\newtheorem{lemma}{Lemma}
\usepackage{comment}
\usepackage[utf8]{inputenc}
\usepackage{amsmath}
\usepackage{float}
\usepackage{mathtools}
\usepackage{braket}
\usepackage{graphicx}
\graphicspath{{./figures/}}
\usepackage{amsfonts}

\usepackage[ruled,vlined,linesnumbered]{algorithm2e}
\let\oldnl\nl
\newcommand{\nonl}{\renewcommand{\nl}{\let\nl\oldnl}}

\usepackage{csquotes}
\usepackage{hhline}
\usepackage{amssymb}
\usepackage{listings}
\usepackage{color}
\definecolor{codegreen}{rgb}{0,0.6,0}
\definecolor{codegray}{rgb}{0.5,0.5,0.5}
\definecolor{codepurple}{rgb}{0.58,0,0.82}
\definecolor{backcolour}{rgb}{0.95,0.95,0.92}

\usepackage{subcaption}
\usepackage{titlesec}

\usepackage{dcolumn}
\usepackage{tabularx}
\usepackage{hyperref}
\usepackage{amsmath}
\usepackage{multirow}

\usepackage{algorithmic}
\usepackage{graphicx}

\hypersetup{
    colorlinks=true,
    linktoc=all,
    linkcolor=black,
    citecolor=magenta,
}
\usepackage{longtable}
\usepackage{braket}
\newcolumntype{C}{>{\centering\arraybackslash}X}
\SetAlFnt{\small}
\usepackage[font=small]{caption}
\usepackage{amsmath, amssymb, amsthm}
\newtheorem{theorem}{Theorem}

\begin{document}
\hyphenation{op-tical net-works semi-conduc-tor}

\title{Search-for-Gradient: Search-Based Restarts for Optimisation in Structured Flat Landscapes of Variational Quantum Algorithms.
}


\author{Subham Das\thanks{Subham Das is with the Department of Instrumentation and Applied Physics, Indian Institute of Science, Bengaluru, Karnataka 560012, India, email:(subham1@iisc.ac.in).}, Bikash~K.~Behera\thanks{B.~K. Behera is with Bikash's Quantum (OPC) Pvt. Ltd., Mohanpur, WB, 741246 India, and Università degli Studi di Cagliari, Via Is Mirrions, Cagliari, 09123, Italy, e-mail: (bikas.riki@gmail.com).}, Andrés Camilo Granda Arango\thanks{A.~C.~Granda Arango is with Università degli Studi di Cagliari, Via Is Mirrions, Cagliari, 09123, Italy, e-mail: (a.grandaarango@gmail.com).}, Giuseppe Sergioli\thanks{G. Sergioli is with Università degli Studi di Cagliari, Via Is Mirrions, Cagliari, 09123, Italy, e-mail: (giuseppe.sergioli@gmail.com).}, and Roberto Giuntini\thanks{R. Giuntini is with Università degli Studi di Cagliari, Via Is Mirrions, Cagliari, 09123, Italy, and Technische Universität München. Institute for Advanced Study (IAS), Lichtenbergstraße 2a, 85748 Garching b. M\"unchen, Germany, e-mail: (giuntini@unica.it).}}%

\maketitle
\begin{abstract}
Variational Quantum Algorithms (VQAs) are a promising class of hybrid quantum-classical algorithms that involve parameterised quantum circuits and classical optimisation. VQAs have found applications in simulation, optimisation, and machine learning. However, their scalability is hindered by flat landscapes—regions in the parameter space where gradients vanish exponentially small. In such regimes, gradient-based optimisation methods fail because parameter updates become ineffective in the absence of informative gradients.
In this work, we propose a search-based restart framework to escape flat regions in structured flat landscapes, in which small trainable regions exist within a large flat landscape. This framework approaches the plateau escape as a search problem and locates trainable regions. The optimiser's parameters stop updating when the gradients vanish, and the method then searches over-sampled points to identify landscapes with nonzero gradients. We provide theoretical estimates of the number of jumps and the probability of escape from flat landscapes. This restarts the local optimisation process and avoids stagnation. We tested this framework on synthetic functions that mimic structured flat landscapes and a real quantum circuit cost landscape. Theoretical analysis and experimental results demonstrate that Search-for-Gradient improves optimisation performance in structured flat landscapes. The proposed framework assumes the existence of a trainable region with nonzero measure and will not work if it has zero measure. Our framework can be integrated with existing VQAs to improve trainability and robustness.
\end{abstract}
\IEEEpeerreviewmaketitle

\section{Introduction}
The rapid development of quantum hardware has driven significant advances in quantum algorithms in the noisy intermediate-scale quantum (NISQ) era, and one important question is whether NISQ-era computers will have practical applications \cite{Preskill_2018}. The NISQ era faces both resource constraints and gate-fidelity issues. Thus, hybrid quantum-classical algorithms constitute a leading class of promising, application-oriented approaches. Variational Quantum Algorithms (VQAs) \cite{Cerezo2021} are a leading class of hybrid quantum–classical algorithms with potential applications in chemistry \cite{mustafa2022variationalquantumalgorithmschemical}, condensed matter physics \cite{PhysRevLett.131.060406}, optimization \cite{PhysRevResearch.7.023141,PhysRevResearch.2.033402}, and quantum machine learning \cite{Park2023, PRXQuantum.4.010328}. These algorithms use parameterised quantum circuits and classical optimisers to solve problems in different domains, such as quantum variational eigensolver \cite{McClean_2016, Peruzzo2014, Yung2014}, medical and healthcare systems \cite{11071993}, portfolio optimization\cite{Wang2025}, drug discovery \cite{Li2024}, cryptography\cite{YANG2025131474}, etc. 

Despite its potential applications, a major obstacle to this optimisation is the presence of flat loss landscapes, commonly referred to as barren plateaus, where the gradients of the objective function become exponentially small as the size increases. The barren plateau was first introduced in the work \cite{mcclean2018barren}. McMcClean et al. showed that any randomly initialised parameterised quantum circuit tends to approximate unitary 2-designs, leading to the expectation of the gradient being zero and the gradient variance of the cost function vanishing exponentially as the number of qubits increases. Consequently, this makes gradient estimation prohibitively expensive and renders gradient-based optimisation ineffective for large-scale variational circuits. The work \cite{Cerezo2021} showed that the local gradient of a variational circuit largely depends on the cost function. It was shown that global cost functions exhibit vanishing gradient variance exponentially, even for shallow circuits.

In contrast, local cost functions maintain gradients that decrease at worst polynomially with the number of qubits, provided the circuit depth scales as $O(\log(n))$. Moreover, the work shows that \cite{PhysRevA.111.012441} found that barren plateau regions can be populated by exponentially many low-quality local minima, in which only a small subset of the objective function is optimised. At the same time, the remaining terms remain on barren plateaus.

To address the barren plateau problem, several methods have been proposed. Existing methods include parameter initialisation strategies \cite {PhysRevA.106.042433, Grant2019initialization}, shallow trainable circuits and local cost function \cite{cerezo2021cost}, layer-wise Optimizers\cite{Skolik2021} and restart Strategies that reinitialise the optimization when progress halts \cite{liu2025escapingbarrenplateaucoexploration}. However, these methods primarily rely on local gradient information. When trainable regions are sparse and embedded within a vast flat landscape, local methods cannot find them. Even the repeated restarts method may waste significant resources inefficiently exploring flat regions before encountering informative gradients.

In this work, we propose a search-based restart mechanism to identify such regions in a structured landscape, where local gradient descent fails. However, trainable regions with non-zero gradients exist elsewhere in the parameter space. In this work, we reframe the gradient vanishing problem as a search problem over the parameter space. This allows us to identify new starting points with non-vanishing gradients, enabling continued optimisation. Our framework is compatible with a wide range of search and sampling strategies, and can be integrated with existing VQAs to improve trainability and robustness. The structured landscapes are particularly well-suited for adaptive search-based optimisation, and we provide both theoretical and empirical evidence to support this claim.

\section{Motivation and Related Work}
Flat landscapes have been a major challenge in optimising Quantum Neural Networks (QNNs). According to work \cite{cerezo2021cost}, for a shallow circuit, the variance of the gradients depends on the cost function. For a Global cost function, gradient variance vanishes exponentially with qubit count n, even if circuit depth is $O(\log n)$. For the local cost function, the gradient variance vanishes at worst polynomially with n, provided circuit depth is $O(\log n)$. Thus, a large set of circuits whose depth is beyond  $O(\log n)$ gets trapped in the barren plateau problem, and thus training those circuits and architectures becomes an extremely hard problem. In the work \cite{Ragone2024}, a unified Lie algebraic framework for understanding the barren plateau is presented. This problem has been extensively studied \cite{mcclean2018barren}, with causes linked to circuit expressibility, entanglement, and noise. Prior work has explored initialisation strategies \cite{PhysRevA.106.042433}, where classical neural networks were used to generate the initial parameters of the circuit \cite{PhysRevA.106.042433}. In the work \cite{Grant2019initialization}, a different initialisation strategy was proposed in which, instead of randomly initialising all circuit parameters, a subset of parameters is chosen so that the parameterised circuit is initially composed of shallow blocks, effectively reducing the circuit depth during the initial optimisation stage. In the work layer-wise training \cite{skolik2021layerwise}, the circuit depth is gradually increased during optimisation.
In contrast, only a subset of circuit parameters is updated at each training stage, by maintaining a shallow effective circuit depth and optimising fewer parameters at a time. In the work \cite{rad2022survivingbarrenplateauvariational}, a fast-and-slow optimisation algorithm was introduced, which combines Bayesian learning with gradient-based optimisation. In their approach, Bayesian learning is first employed to identify a promising region of the parameter space, after which a local gradient-based optimiser is initialised to refine the solution efficiently. Another work \cite{Arrasmith_2021} shows that using gradient-free optimisation does not inherently avoid the barren plateau problem, as the cost-function differences used to guide the optimisation become exponentially suppressed with the number of qubits. Another interesting work \cite{11389907} proposes a parallel optimisation approach that mitigates barren plateau-induced stagnation by evolving multiple optimisation trajectories simultaneously. Upon hitting a low gradient, the trajectory is reinitialised.

Thus, we propose a novel method of escaping the flat landscape problem. In this method, if the optimiser gets trapped in a flat region, it enters a search phase to find a trainable region in the landscape. Once the trainable region is located, the algorithm resumes gradient-based optimisation. Restart strategies inform our approach to classical optimisation. The method is shown to be effective in the synthetic landscape and a real quantum circuit cost function.

\section{Preliminaries}

We model the parameter space of a variational quantum algorithm as a probability space $(\Theta, \mathcal{F}, \mu)$, where $\Theta$ is the set of all possible parameter vectors, $\mathcal{F}$ is a collection of subsets of $\Theta$ (a sigma-algebra), and $\mu$ is a probability measure with $\mu(\Theta) = 1$.

\subsection{Flat landscape as Measurable Sets}
A flat landscape is a region of the parameter space where the gradient of the cost function is near zero:
\begin{equation}
B := \{\theta \in \Theta \,|\, \|\nabla C(\theta)\| < \epsilon_g \},  B := \Theta \setminus T
\end{equation}
The measure of the flat landscape, $\mu(B) = p_{\rm total} \in (0,1)$, is the fraction of the parameter space occupied by barren regions. In general, there may be multiple disjoint barren regions $B_1, B_2, \dots, B_n$, whose measures sum to $p_{\rm total}$.
\subsection{Trainable region}
A trainable part is a region of the parameter space where the gradient of the cost function is non-zero:
\begin{equation}
T := \{\theta \in \Theta \,|\, \|\nabla C(\theta)\| \ge \epsilon_g \},  T := \Theta \setminus B
\end{equation}regions
Similarly, the trainable region can have multiple regions in space $T_1, T_2, \dots, T_n$. 
\subsection{ Structured flat landscape}

Let \( \Theta \subset \mathbb{R}^n \) be the parameter space of a variational quantum circuit, and let \( C: \Theta \to \mathbb{R} \) be a differentiable cost function. Since, \( \Theta \subset \mathbb{R}^n \) is defined by radians, the $\Theta$ is bounded by $[0,2\pi]^n$. 

We say that \( C \) exhibits a \textbf{structured flat landscape} if:

\textbf{Partitioned Gradient Landscape:} There exists a measurable disjoint partition of the parameter space:
\begin{equation}
\Theta = \bigcup_{i=1}^{M} B_{i} \cup \bigcup_{k=1}^{N} T_k
\label{structuredbarren}
\end{equation}

Here, $B_{i}$ and $T_{k}$ are the regions of flat landscape and trainable region, respectively.

\subsection{Random Jumps}
A random jump corresponds to sampling a point $\theta_i \in \Theta$ according to the uniform measure $\mu$. For a sequence of jumps $\theta_1, \dots, \theta_k$, we define events $E_i = \{\theta_i \in B\}$ to track whether a jump lands in a flat landscape.

\subsection{Escape Probability}
The escape probability after $k$ jumps is the probability of leaving the flat landscape at least once:
\[
P_{\rm escape}(k) := 1 - P\Big(\bigcap_{i=1}^k E_i\Big).
\]
Using this formalism, we can rigorously analyse how random jumps increase the likelihood of escaping a flat landscape in variational algorithms.

\section{Methodology}

\subsection{Sigma Algebra for Structured flat landscape}

The structured, flat landscape provides a natural partition of the parameter space, making it suitable for probabilistic modelling. To construct a probabilistic framework, we first need to define a $\sigma$-algebra on the parameter space so that a measure (probability measure) can be defined on measurable subsets.

The structured flat landscape decomposes the parameter space into disjoint barren regions and trainable regions. Since these regions form a partition of the parameter space, we work with a $\sigma$-algebra generated by flat and trainable regions. 

\begin{equation}
\mathcal{F} = \sigma \left( \left\{ B_j \right\}_{j=1}^{M} \cup \left\{ T_k \right\}_{k=1}^{N} \right)
\end{equation}

This allows us to define measurable events corresponding to the optimiser being located in specific barren regions, trainable regions, or their unions. The formal proof of the $\sigma$-algebra for the structured flat landscape is given in Theorem~\ref{sigma}. Once the $\sigma$-algebra is defined, we can introduce a measure on this measurable space and develop a probabilistic model for the optimisation dynamics over the structured flat landscape.

Having constructed the $\sigma$-algebra $\mathcal{F}$ on $\Theta$, the measurable space $(\Theta, \mathcal{F})$ with a probability measure $\mu$, the $(\Theta, \mathcal{F}, \mu)$ becomes a probability measure space. The parameter space is given as:

\medskip

\begin{equation}
\Theta = [0, 2\pi]^d,
\end{equation}
where $d$ is the number of trainable parameters. Let $\lambda$ denote the Lebesgue measure on $\mathbb{R}^d$. We define the normalised measure
\begin{equation}
\mu(A) = \frac{\lambda(A)}{\lambda(\Theta)}, \quad A \in \mathcal{F}.
\end{equation}

\subsection{Algorithm Overview}

In this section, we describe the overall workflow of the 
\textit{Search-for-Gradient} optimisation framework. The method alternates 
between classical gradient-based optimisation and a search-based restart 
mechanism when the optimiser becomes trapped in a flat landscape.

\subsubsection{Core Workflow}

The algorithm operates in iterative phases consisting of gradient-based 
optimisation, plateau detection, search-based restart, and archive management.

\textbf{Step 1: Classical Optimisation Phase}  
Starting from an initial parameter vector $\theta_0$, a classical 
gradient-based optimiser (e.g., SGD, Adam, or gradient descent) is used to 
minimize the cost function $C(\theta)$. The optimiser continues until either 
Convergence or plateau detection occurs. A flat region is detected when the 
gradient norm satisfies
\begin{equation}
\|\nabla C(\theta)\| < \epsilon
\end{equation}
for $W$ consecutive iterations. The mean and the standard deviation of the cost satisfy the following conditions.
\begin{align}
    \text{mean}>\zeta\\
    \text{standard deviation}< \sigma
\end{align}

\textbf{Step 2: Search Phase}  
Once a plateau is detected, the algorithm enters a one-time search phase. The algorithm searches the parameter space to locate a new point with a higher gradient magnitude. The search procedure consists of the following steps:
\begin{itemize}
\item Estimate the fraction of the trust region occupied by the flat landscape 
regions, denoted by $p_{\text{total-ball}}$ in a given ball of radius $r$.
\item Compute the number of candidate samples required to achieve a target escape probability $P_{\text{target-ball}}$ under the independent sampling assumption:
\begin{equation}
S = \left\lceil 
\frac{\ln(1 - P_{\text{target- ball}})}{\ln(p_{\text{total- ball}})} 
\right\rceil.
\end{equation}
The derivation of this formula is discussed below.
\item Sample $S$ candidate parameter vectors from the parameter space using 
a chosen sampling strategy (Gaussian, Cauchy, Sobol, etc.).
\item Evaluate each candidate using a scoring function such as gradient 
magnitude, cost decrease, or surrogate evaluation.
\item Select the best candidate $\theta^*$ based on the scoring function. If it fails to find a suitable parameter, increase the ball's radius and repeat the process.
\item Once the best parameter is obtained, exit the process.
\end{itemize}

\textbf{Step 3: Restart Phase}  
The optimiser is restarted from the selected candidate point $\theta^*$, and 
gradient-based optimisation resumes from this new initialisation.

\textbf{Step 4: Archive Management}  
An archive of previously visited points is maintained to prevent the optimiser 
from repeatedly visiting the same barren regions. If a newly sampled point is 
within a small Euclidean distance of any archived point, the candidate is 
discarded, and the new best point from the sampled point is chosen.

\textbf{Step 5: Iteration}  
The algorithm alternates between gradient-based optimisation and search-based 
restarts until a stopping criterion is met, such as maximum iterations, target 
cost value, or computational budget.

\textbf{Step 6: Output}  
The algorithm returns the parameter vector with the lowest cost encountered during the optimisation process.

\begin{algorithm}
\caption{Adaptive Single-Shot Search-for-Gradient with Expanding Radius}
\label{alg:sfg}
\DontPrintSemicolon
\DontPrintSemicolon

\KwIn{Initial parameters $\theta_0$, threshold $\epsilon$, plateau window $L$, initial radius $r_0$, increment $\delta$, maximum radius $r_{\max}$, neighborhood radius $\epsilon_{\text{ball}}$, maximum number of jumps $K_{\max}$, maximum iterations $T$, mean threshold $\zeta$, standard deviation threshold $\sigma$, Sample size $S$}

\KwOut{Optimized parameter $\theta$}

$\theta \leftarrow \theta_0$ \\
$r \leftarrow r_0$ \\
$\mathcal{A} \leftarrow \emptyset$  \\ 
$\text{visited} \leftarrow \text{False}$ 

\For{$t = 1$ \KwTo $T$}{

Perform gradient-based optimisation step\;
\If{\textbf{not} visited \textbf{and} 
$\|\nabla C(\theta)\| < \epsilon$ for $L$ steps \textbf{and}
$\text{mean}(C_{t-L:t}) > \zeta$ \textbf{and}
$\text{std}(C_{t-L:t}) < \sigma$}{

$K \leftarrow 0$\;

\While{$K < K_{\max}$}{

    Sample $\{\theta'_i\}_{i=1}^S$ within radius $r$\;
    Compute $G(\theta'_i) = \|\nabla C(\theta'_i)\|$\;

    $\theta^* \leftarrow \arg\max_{\theta'_i} G(\theta'_i)$\;

    \While{$\theta^*$ is within $\epsilon_{\text{ball}}$ of any point in $\mathcal{A}$}{
        Select the next best candidate\;
    }

    \If{$G(\theta^*) > \epsilon$}{
        $\theta \leftarrow \theta^*$\;
        $\mathcal{A} \leftarrow \mathcal{A} \cup \{\theta^*\}$\;
        $r \leftarrow r_0$\;
        \textbf{break}\;
    }
    
    \Else{
        $r\leftarrow\min(r+\delta,r_{\max})$\;
        $K\leftarrow K+1$\;
    }
}

}
visited $\leftarrow$ True\;

}

\Return $\theta$

\end{algorithm}

\subsection{Restarts}
In the algorithm, three different types of restarts were used. These restarts are as follows:
\subsubsection{Independent global restarts} 
A sequence of jumps $\{\theta_1, \theta_2, \dots, \theta_k\}$ is said to be \emph{independent} if the probability of landing in a flat landscape does not depend on previous jumps, i.e.,
\begin{equation}
P(\theta_i \in B \mid \theta_1, \dots, \theta_{i-1}) = P(\theta_i \in B) = p_{\rm total}, \quad \forall i,
\end{equation}
where $B \subset \Theta$ denotes the flat landscape region and $p_{\rm total} = \mu(B)$ is its measure.

\subsubsection{local adaptive restarts}
A sequence of jumps $\{\theta_1, \theta_2, \dots, \theta_k\}$ is said to be a \emph{local adaptive restart} if the probability of landing in a region depends on the currently landed region, i.e.,
\begin{equation}
P(\theta_i \in B \mid \theta_1, \dots, \theta_{i-1}) = P(\theta_i \in B\mid \theta_1) = p_{\rm adaptive}, \quad \forall i,
\end{equation}
\subsubsection{archive-aware restarts}
A sequence of jumps $\{\theta_1, \theta_2, \dots, \theta_k\}$ is said to be an \emph{archive-aware restart} if the probability of landing in a region depends on the previously landed region, i.e.,
\begin{equation}
P(\theta_i \in B \mid \theta_1, \dots, \theta_{i-1}) = p_i(B),
\end{equation}
where $p_i(B)$ is an adaptive probability that depends on the history $\{\theta_1, \dots, \theta_{i-1}\}$, and $B \subset \Theta$ denotes the flat landscape region.

\begin{lemma}[Escape probability from flat landscape for independent jumps]
    
Let $(\Theta, \mathcal{F}, \mu)$ be the parameter space with $\mu(\Theta) = 1$. 
Suppose the space contains $n$ barren regions $B_1, \dots, B_n \subset \Theta$ 
Then, the mean number of independent jumps to escape the flat landscape is
\begin{equation}
\mathbb{E}[K] = \frac{1}{1 - p_{\text{total}}}.
\label{expectedjump}
\end{equation}

\end{lemma}
\begin{proof} Suppose for each jump the algorithm succeeds in escaping the flat landscape with probability 
\begin{equation}
    p_{train}=q=1-p_{\rm total}
\end{equation}

The probability that after failing $k-1$ jumps, one succeeds in escaping the flat landscape in $k$ jumps is given by

\begin{equation}
    P(K=k)=(p_{total})^{k-1}q
\end{equation}

Thus, the mean number of trials for escaping the flat landscape for an independent jump is 

\begin{equation}
\mathbb{E}[K] = \frac{1}{1 - p_{\text{total}}}=\frac{1}{p_{train}}
\end{equation}
\label{thm 1}
\end{proof}

\subsection{Sampling technique}
The expectation value in Eq.~\ref{thm 1} is derived under the assumption of independent sampling. If the sampling technique is not independent, this expression is invalid. However, this expression can be used to estimate the number of samples. 
\subsubsection{Sampling distributions}
We consider three sampling strategies in SFG: Gaussian, Cauchy, and Sobol sampling. Each strategy provides a different point distribution and, hence, a different exploration behaviour.

\begin{itemize}
    
\item \text{Gaussian sampling:}
In Gaussian sampling, the random points are sampled from an isotropic Gaussian distribution. For a Gaussian distribution, we can expect the sampled points to be within the $S\sim (2-3) \sigma$ range of the Gaussian distribution. Here S represents the sampled points and $\bar{S}$ represents the number of sampled points. Thus, it is perfect for sampling when we suspect the flat region that our optimiser is escaping from is close to the current point. Since the sampling is in proximity, the expected average number of jumps will increase. Thus, we have to sample more points than in the independent case. The Gaussian probability distribution function is 
\begin{equation}
p(x) = \frac{1}{\sqrt{2\pi\sigma^2}} 
\exp\left(-\frac{(x - \mu)^2}{2\sigma^2}\right)
\end{equation}

\item \text{Cauchy Sampling:}
In Cauchy sampling, the random points are sampled from this distribution. The sample mean for the Cauchy distribution does not converge, regardless of the number of samples, so the Sample variance does not exist. Calculating it from samples gives wildly fluctuating results. Thus, outliers are common because the probability of large deviations is significant. Hence, it is particularly useful for global escape or for sampling points outside the centre of the probability distribution function.
\begin{equation}
f(x; x_0, \gamma) = \frac{1}{\pi \gamma} \cdot 
\frac{\gamma^2}{(x - x_0)^2 + \gamma^2}, \quad \gamma > 0
\end{equation}

\item \text{Sobol sampling:}
In Sobol sampling, the points are selected uniformly. For our SFG, we implemented the unscrambled Sobol sampling \cite{SOBOL196786}. It is a deterministic sampling method designed to cover the parameter space uniformly with minimal clustering. The uniform coverage of the parameter space reduces the likelihood of redundant sampling. The method uses samples in powers of $2^m$. For SFG, when sample numbers are not powers of 2, we use the sample number closest to a power of 2. 

\end{itemize}

\subsubsection{Sampling radius}

Sampling in SFG is performed within a controlling parameter, which we call the radius. It does not mean the sampled point will be within the radius of the centre; rather, it is a hyperparameter that controls the distribution's spread. For a given search radius $r$, candidate points are generated by sampling
displacement vectors from the selected sampling distribution. Let $x$ denote the current point in the parameter space and let $S$ denote a sample. A candidate point $x_{\mathrm{new}}$ is then constructed
as
\begin{equation}
    x_{\mathrm{new}} = x + rS.
\end{equation}
Here, the parameter $r$ controls the scale of the search region, while the sampling distribution determines the spatial distribution of the candidate points around the current point. Since the Sobol sampling is carried out on $[0,1]^n$, this transformation will not be uniformly spread. Thus, for Sobol sampling, first a transformation is carried out for each coordinate, where the sampled points $u_{ij}$ are transformed from the domain $[0,1]$ to $[-1,1]$ using the transformation
\begin{equation}
    u \longrightarrow (2u_{ij} - 1)
\end{equation}

\begin{figure}
    \centering
    
    \begin{subfigure}[b]{0.3\linewidth}
        \centering
        \includegraphics[width=\linewidth]{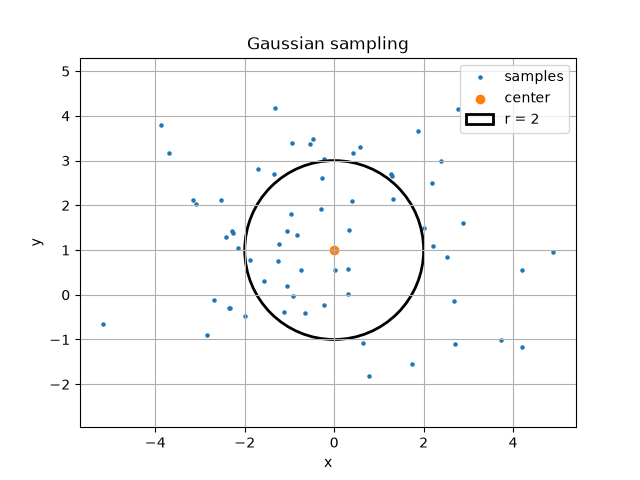}
        \caption{Gaussian sampling}
        \label{samplingGaussian}
    \end{subfigure}
    \hfill
    \begin{subfigure}[b]{0.3\linewidth}
        \centering
        \includegraphics[width=\linewidth]{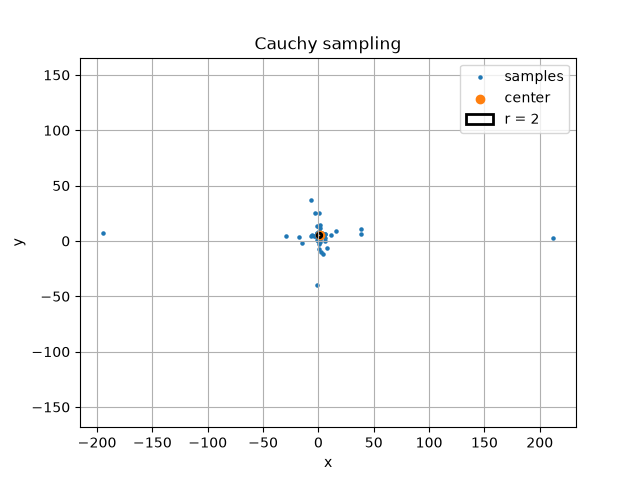}
        \caption{Cauchy Sampling}
        \label{samplingCauchy}
    \end{subfigure}
    \hfill
    \begin{subfigure}[b]{0.3\linewidth}
        \centering
        \includegraphics[width=\linewidth]{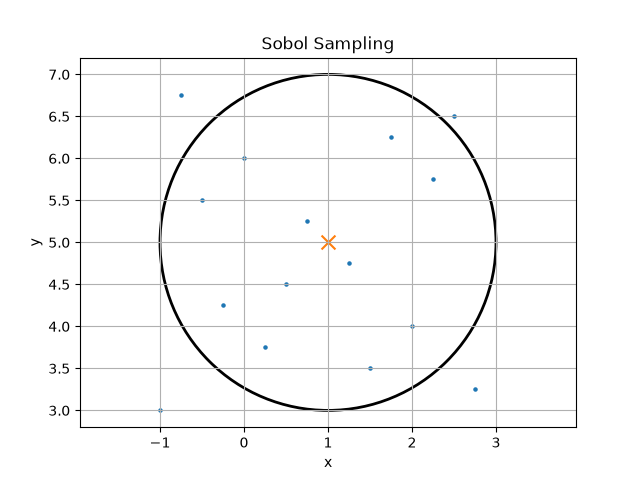}
        \caption{Sobol Sampling}
        \label{sobolbounded}
    \end{subfigure}

    \begin{subfigure}[b]{0.6\linewidth}
        \centering
        \includegraphics[width=\linewidth]{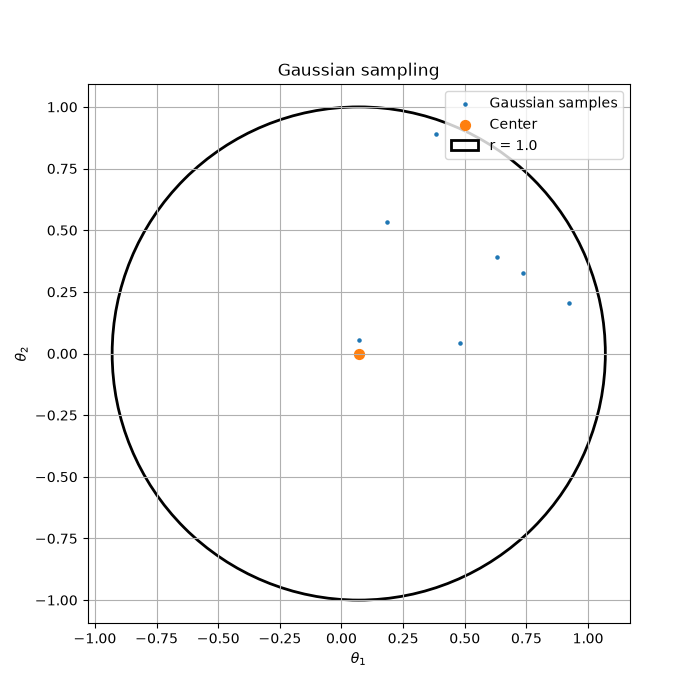}
        \caption{Gaussian sampling in a bounded domain}
        \label{samplingGaussianbounded}
    \end{subfigure}

    \hfill

    \caption{
Comparison of the spatial distributions generated by the different unbounded sampling
strategies for a sampling radius of $r=2$. The bounded Gaussian sampling radius uses $r=1$. 
(a) Gaussian sampling produces a concentrated distribution around the sampling
centre. 
(b) Cauchy sampling exhibits substantially heavier tails, generating occasional
samples far from the centre. 
(c) Sobol sampling provides a deterministic, quasi-uniform coverage of the
sampling region. 
(d) Gaussian sampling restricted to the bounded parameter domain, illustrating the rejection of samples outside
the allowed parameter space.
}
    \label{fig:sampling}
\end{figure}

This transformation produces candidate points within a hypercube centred at x, with side length 2r. The low-discrepancy structure of the Sobol sequence is preserved under this affine transformation, allowing the candidate points to explore the search region systematically. The resulting spatial distributions for Gaussian, Cauchy, and Sobol sampling are illustrated in subfigures~\ref{samplingGaussian},
\ref{samplingCauchy}, and~\ref{sobolbounded}, respectively. Gaussian sampling
generates points concentrated around the current point, whereas the
heavy-tailed Cauchy distribution allows occasional long-range candidate
points. Sobol sampling provides a deterministic, quasi-uniform distribution
of candidate points over the search region.

When the parameter space is bounded, candidate points falling outside the
allowed domain are discarded. In particular, for the bounded domain,
only points satisfying the domain constraints are retained, as illustrated
in subfigure~\ref{samplingGaussianbounded}.

\subsection{Escape probability and Complexity analysis}
\subsubsection{Escape probability}
Given a flat, structured landscape, the escape probability will differ across sampling methods. Under independent random sampling, the probability of observing at least one sample outside the flat region after k jumps is given by the following lemma  \ref{thm2}.

\begin{lemma}[Escape Probability from flat landscape ]
Let $(\Theta, \mathcal{F}, \mu)$ be the parameter space with $\mu(\Theta) = 1$. 
Suppose the space contains $n$ flat regions $B_1, \dots, B_n \subset \Theta$  
with total flat region fraction
\begin{equation}
p_{\rm total} := \mu\Big(\bigcup_{i=1}^n B_i\Big) \in (0,1).
\end{equation}

Consider an algorithm performing $k$ random independent jumps $\theta_1, \dots, \theta_k$.

Then, the escape probability satisfies.
\begin{equation}
P_{\rm escape}(k) = 1 - P(\theta_1, \dots, \theta_k \in B) = 1 - p_{\rm total}^k 
\end{equation}

\begin{proof}
Let $E_i := \{\theta_i \in B\}$ denote the event that the $i$-th jump lands in a flat region 
where $B := \bigcup_{i=1}^n B_i$.

\textbf{ Single jump.} By definition of measure:
\begin{equation}
P(E_1) = \mu(B) = p_{\rm total}.
\end{equation}

 Since the jumps are independent:
\begin{equation}
P(E_i \mid \theta_1, \dots, \theta_{i-1}) =  P(\theta_i \in B) = p_{\rm total}, \quad \forall i.
\end{equation}

\textbf{Joint probability of being trapped.} By the chain rule of probability:
\begin{align*}
P(E_1 \cap E_2 \cap \dots \cap E_k) 
&=  \prod_{i=1}^k P(E_i) = p_{\rm total}^k.
\end{align*}

\textbf{Escape probability:} 
\begin{equation}
P_{\rm escape}(k) = 1 - p_{\rm total}^k.
\label{escape}
\end{equation}

\end{proof}
\label{thm2}
\end{lemma}

\subsubsection{Time Complexity}

Four main factors govern the complexity of the proposed method: 
(i) the cost of generating a sample according to the chosen sampling strategy, 
(ii) the number of samples $S$ drawn in each search round, 
(iii) the cost of evaluating the gradient of the cost function at each sampled point, and 
(iv) the archive proximity test used to avoid redundant restarts. 
For Sobol sampling in $d$ dimensions, generating a single candidate point requires $O(d)$ operations through XOR updates with precomputed direction numbers. Evaluating the gradient at that point costs $O(C_{eval}\theta)$, where $\theta$ denotes the number of trainable parameters. 
Thus, drawing and evaluating $S$ candidates requires $O(S(d+C_{eval}\theta))$ time. 
After selecting the best candidate, the algorithm checks whether it lies within a $\tau$--ball of any previously visited point. 
This requires computing a squared Euclidean distance in $\mathbb{R}^d$ against each of the $M$ archived points, which costs $O(d)$ per comparison and therefore $O(Md)$ in total. Hence, one complete search round has time complexity
\begin{equation}
    O\big(S(d+ C_{eval}\theta) + Md\big).
\end{equation}
The algorithm performs at most $K_{\max}$ search rounds after plateau detection. A good estimate of the total number $K$ required to escape the flat region can be obtained from Lemma \ref{thm 1}, giving $K_{expected}=\frac{1}{p_{train}}$. However, the worst-case possibility is that it exhausts the whole $K_{\max}$. Thus, the  number of jumps is still $K_{max}$, giving the total worst-case time complexity to be:
\begin{equation}
O\big(K_{\max}(S(d+C_{eval}\theta) + Md)\big).
\end{equation}
Noting that at most one point is added to the archive per search round, we have $M \leq K_{\max}$ in the worst case. 
Therefore, the bound is tight and directly reflects the sampling budget, the dimensionality of the search space, and the number of restarts required to escape a structured flat landscape. A similar analysis can be performed for other sampling techniques; the time complexity of each method is shown in Table \ref{tab:complexity_linear}.

\begin{table*}
\renewcommand{\arraystretch}{0.5}
\caption{Time complexity of the complete search phase using linear archive lookup.
Here $S$ is the number of sampled candidates per round, $d$ is dimension, $\theta$ is the 
number of parameters for gradient computation, $M$ is the archive size, and $K_{\max}$ is
the maximum number of search rounds.}
\centering
\begin{tabular}{|c|c|c|c|}
\hline
\textbf{Component} & \textbf{Time complexity}  \\ \hline
Sampling per candidate   & $O(d)$            \\ \hline
Gradient per candidate   & $O(C_{circuit}\theta)$   \\ \hline
Selecting $S$ candidates  & $O(S(d+C_{circuit}\theta))$             \\ \hline
Archive proximity check  & $O(Md)$           \\ \hline
\textbf{Total (one round)} & $O(S(d+C_{circuit}\theta) + Md)$ \\ \hline
\textbf{Total (full search phase)} & $O(K_{\max}(S(d+C_{circuit}\theta) + Md))$ \\ \hline
\end{tabular}
\label{tab:complexity_linear}
\end{table*}

\subsubsection{Space complexity}
The method's space usage comes from three components: the sampling mechanism, the archive, and a small number of working buffers. The additional space overhead for Sobol sampling comes from storing the direction numbers used to generate low-discrepancy points. For each dimension, the Sobol generator maintains $L = \lceil \log_2 N_{\max} \rceil$ direction 
numbers, resulting in $O(d \log N_{\max})$ storage. Gaussian and 
Cauchy sampling does not require precomputed tables and therefore uses only a constant sampler state. All methods maintain an archive of previously accepted points, which requires $O(Md)$ memory since each entry is $d$-dimensional and 
at most $M$ such points are stored. During a search round, only the current candidate, the best candidate, and the gradient vector are stored, requiring  $O(d + \theta)$ space. Thus, the overall space complexity becomes 
$O(d \log N_{\max} + Md + \theta)$ for Sobol sampling and $O(Md + \theta)$ 
for Gaussian or Cauchy sampling.

\begin{table*}
\renewcommand{\arraystretch}{0.6}
\caption{Space complexity for different sampling strategies. 
Here $d$ is the dimension, $\theta$ is the number of parameters, 
$M$ is the archive size, and $N_{\max}$ is the maximum number of samples generated.}
\centering
\begin{tabular}{|c|c|c|c|}
\hline
\textbf{Component} & \textbf{Sobol} & \textbf{Gaussian} & \textbf{Cauchy/Levy} \\ \hline

Sampler state  
& $O(d \log N_{\max})$  
& $O(1)$  
& $O(1)$  \\ \hline

Archive storage  
& $O(Md)$  
& $O(Md)$  
& $O(Md)$ \\ \hline

Working buffers  
& $O(d + \theta)$  
& $O(d + \theta)$  
& $O(d + \theta)$  \\ \hline

\textbf{Total}  
& $O(d \log N_{\max} + Md + \theta)$  
& $O(Md + \theta)$  
& $O(Md + \theta)$  \\ \hline
\end{tabular}
\label{tab:space_complexity}
\end{table*}

\subsection{Boundary Condition}
\subsubsection{Maximum Restarts ($K_{\max}$)}
The parameter $K_{\max}$ sets an upper limit on the maximum number of search rounds within the single search phase. A larger $K_{\max}$ increases the chance of finding a global optimum but also increases computational cost. If the algorithm escapes plateaus efficiently, $K_{\max}$ can be smaller; otherwise, it should be increased to allow sufficient exploration. $P_{\mathrm{target}}$ is the probability with which it wants to escape the flat region. Under independent sampling, the required number of candidate draw events can be estimated using equation \ref{escape}, which gives:
\begin{equation}
K_{target}
=
\left\lceil
\frac{\log\left(1-P_{\mathrm{target}}\right)}
{\log\left(p_{\mathrm{total}}\right)}
\right\rceil.
\label{k_max}
\end{equation}
Under independent sampling with a fixed flat-region probability $p_{\mathrm{total}}$, the random number $K$ of candidate draws required for the first successful escape satisfies equation \ref{thm 1}. This expectation value should be distinguished from the deterministic budget $K_{\mathrm{max}}$. In Algorithm~\ref{alg:sfg}, $K_{\max}$ limits search rounds, each of which may evaluate multiple candidates; accordingly, the formula serves as a reference for budget selection, while a round-level guarantee depends on the sampling and candidate-selection assumptions.
\subsubsection{Number of Candidate Samples per Plateau ($S$)}
The parameter $S$ determines the number of candidate points sampled during each plateau-escape search. $S$ should be chosen to ensure a high probability of escaping a barren region in a single iteration. Let $p_{\rm total-ball}$ denote the fraction of the trust region covered by the flat region. Then, for a target escape probability $P_{\rm target-ball}$, we can estimate the number of samples:
\begin{equation}
S = \left\lceil \frac{\ln(1 - P_{\rm target-ball})}{\ln(p_{\rm total-ball})} \right\rceil
\label{sample_r}
\end{equation}
The value of $S$ can be adjusted depending on the sampling strategy. If $P_{target}$ and $P_{total}$ are same as $P_{\rm target-ball}$ and $p_{\rm total-ball}$ respectively. The equations \ref{k_max} and \ref{sample_r} coincide only if a round contains exactly one candidate.
\subsubsection{Distinguishing Candidate Draws from Search Rounds}
Under the additional assumptions that each round tests exactly S independent candidates, all candidates share the same failure probability $P_{total}$, and none is rejected, the chance of finding a trainable point within R rounds is
\begin{align}
\Pr(\text{success within } R \text{ rounds})
= 1 - p_{\mathrm{total}}^{SR}
\end{align}
To achieve a target escape probability \(P_{\mathrm{target}}\), the required number of search rounds is
\begin{equation}
R_{\mathrm{target}}=\left\lceil
\frac{\log(1-P_{\mathrm{target}})}
{S\log(p_{\mathrm{total}})}
\right\rceil.
\label{eq:rounds_target}
\end{equation}
Here, $R_{target}$ is the minimum number of search rounds required to reach a desired probability of escaping a flat region. Here, \(R_{\mathrm{target}}\) is the minimum number of search rounds required to reach the desired probability of escaping a flat region. The total number of candidate draws is \(N=SR\) under these assumptions.

Equation \ref{eq:rounds_target} is not a general guarantee for SFG: the radius changes, Gaussian and Cauchy sampling are not uniform on $\Theta$, the Sobol sequence is deterministic, and domain/archive rejection may alter the candidate distribution. In those cases, the simple power law requires appropriate modifications. In particular, a probability estimated under uniform sampling on a ball cannot automatically be used as the failure probability of a non-uniform sampling scheme.

For clarity, \(N\) denotes the total number of candidate draws, \(R\) denotes the number of search rounds, and \(K_{\max}\) denotes the implemented upper limit on the number of search rounds. These quantities are distinct: each round may involve multiple candidate draws, whereas \(K_{\max}\) limits the number of rounds rather than the total number of candidates evaluated.

\subsubsection{Stopping Criteria}
The whole optimisation process is given a fixed number of iterations. The optimisation terminates when the maximum number of iterations (T) is reached. Each step of gradient descent counts as one iteration, and each candidate evaluation in search rounds contributes to a single iteration. In each round, S candidates are drawn. Since (S) candidates are evaluated in each search round, a complete search round consumes (S) iterations. Once the plateau is detected, SFG is activated. The total number of candidate draw rounds is at most $K_{max}$; the optimiser exits the search phase early upon finding a candidate. If no acceptable candidate is found after $K_{\max}$ rounds, the search procedure is abandoned, and gradient descent continues from the current parameter point. The budget is shared between gradient-descent updates and candidate evaluations. The optimisation terminates when the cumulative iteration count reaches T.

\section{Analytical examples}
In this section, we present a few analytical examples to illustrate the model's usefulness. 
\subsection{ Coherent State Variational Circuit}

Consider a variational quantum circuit that prepares a coherent state using the displacement operator
\begin{equation}
U(\alpha) = e^{\alpha a^\dagger - \alpha^* a},
\end{equation}
where $\alpha = x + iy$ is a complex parameter such that $x,y\in[-\pi,\pi]$. We define the cost function as the overlap with the vacuum state,
\begin{equation}
C(x,y) = 1 - |\bra{0}U(\alpha)\ket{0}|^2.
\end{equation}
Using the known overlap relation for coherent states,
\begin{equation}
\bra{0}U(\alpha)\ket{0} = e^{-\frac{|\alpha|^2}{2}},
\end{equation}
The cost function becomes
\begin{equation}
C(x,y) = 1 - e^{-(x^2 + y^2)}.
\end{equation}

The gradient of the cost function is
\begin{equation}
\frac{\partial C}{\partial x} = 2x e^{-(x^2 + y^2)},
\end{equation}
\begin{equation}
\frac{\partial C}{\partial y} = 2y e^{-(x^2 + y^2)}
\end{equation}
Thus, the gradient vector can be written as
\begin{equation}
\nabla C = 2e^{-r^2} (x,y),
\end{equation}
where
\begin{equation}
r^2 = x^2 + y^2.
\end{equation}

Using gradient descent with learning rate $\eta$, the update equations are
\begin{equation}
x_{t+1} = x_t - 2\eta  x_t e^{-r_t^2},
\end{equation}
\begin{equation}
y_{t+1} = y_t - 2\eta y_t e^{-r_t^2}.
\end{equation}
This can be written in a factored form as
\begin{equation}
x_{t+1} = x_t \left(1 - 2\eta e^{-r_t^2}\right),
\end{equation}
\begin{equation}
y_{t+1} = y_t \left(1 - 2\eta e^{-r_t^2}\right).
\end{equation}
Hence, the optimisation dynamics are purely radial.

In the continuous gradient flow limit,
\begin{equation}
\dot{x} = -2x e^{-r^2}, \quad
\dot{y} = -2y e^{-r^2}.
\end{equation}
This gives the radial equation
\begin{equation}
\dot{r} = -2r e^{-r^2}.
\end{equation}

Near the minimum ($r \approx 0$),
\begin{equation}
e^{-r^2} \approx 1,
\end{equation}
so
\begin{equation}
\dot{r} \approx -2r,
\end{equation}
which leads to exponential convergence,
\begin{equation}
r(t) = r_0 e^{-2t}.
\end{equation}

However, far from the origin ($r \gg 1$),
\begin{equation}
e^{-r^2} \approx 0,
\end{equation}
and therefore
\begin{equation}
|\nabla C| \approx 0.
\end{equation}
The gradient magnitude is
\begin{equation}
|\nabla C| = 2r e^{-r^2},
\end{equation}
which vanishes exponentially as $r$ grows large. Thus, if the initial parameters are far from the origin, the gradients become extremely small, and the optimisation gets trapped in a flat region of the cost landscape.

In this example, the flat landscape does not arise from circuit depth or randomness, but from the Gaussian decay of the overlap between displaced coherent states in phase space,
\begin{equation}
\bra{0}U(\alpha)\ket{0} = e^{-\frac{|\alpha|^2}{2}}.
\end{equation}
When the parameters are far from the origin or too close to it, the overlap becomes exponentially small, causing the gradient of the cost function to vanish and resulting in a flat landscape in the optimisation space.

A measure of the trainable region can be given through the following analysis. Consider the cost function for the coherent state variational circuit
\begin{equation}
C(x,y) = 1 - e^{-(x^2 + y^2)}.
\end{equation}
The gradient magnitude is given by
\begin{equation}
|\nabla C| = 2r e^{-{r^2}},
\end{equation}

where $( r^2 = x^2 + y^2 )$.

For large r, the exponential term suppresses the gradient, leading to a flat landscape. Therefore, gradient descent optimisation is effective only within a finite annulus where the gradient is non-negligible. We define this region as the \textit{trainable region}. Mathematically, this region can be defined as:
\begin{equation}
  \{ T_{\epsilon_g}={(x,y): 2r e^{-r^2}\ge \epsilon_g}\}
\end{equation}

Now, we can express the inner radius and outer radius of the annulus as

\begin{equation}
r_-=
\sqrt{
-\frac{1}{2}
W_0\left(
-\frac{\varepsilon_g^2}{2}
\right)
},
\qquad
r_+=
\sqrt{
-\frac{1}{2}
W_{-1}\left(
-\frac{\varepsilon_g^2}{2}
\right)
}.
\end{equation}

Thus, gradient descent initialised outside this region will effectively be trapped in a flat landscape, and restart-based methods are required to reach the trainable region. Since we know the radius of the trainable region, the area of the trainable region can be calculated as follows:

\begin{equation}
A_{\text{train}} = \pi\{ (r_{\text{+}})^2-(r_{-})^2\}=\frac{\pi}{2}
\left[
W_0\left(-\frac{\varepsilon_g^2}{2}\right)
-
W_{-1}\left(-\frac{\varepsilon_g^2}{2}\right)
\right].
\end{equation}

The area of the trainable landscape is valid when $r_{out}\leq\pi$. If $r_{out}>\pi$, the annulus extends outside the parameter domain. In that case, we must find the area of intersection between the annulus and $\Theta$ to calculate the trainable region. The parameter space of the variational circuit is bounded by
\begin{equation}
\Theta = [-\pi,\pi] \times [-\pi,\pi],
\end{equation}
So the total parameter space area is
\begin{equation}
A_{\text{total}} = (2\pi)^2 = 4\pi^2.
\end{equation}

Since, the total area is $A_{\text{total}}$, it follows $A_{\text{train}}$ must satisfy the constraint
\begin{equation}
  A_{\text{train}}\leq  A_{\text{total}}
\end{equation}

Assuming uniform sampling over $\Theta=[-\pi,\pi]^2$ and assuming the trainable region is contained entirely within $\Theta$, the probability of landing in the trainable region is the ratio of its area to the total area.
\begin{equation}
p_{\text{train}} =
\frac{A_{\text{train}}}{A_{\text{total}}}
=\frac{\frac{\pi}{2}
\left[
W_0\left(-\frac{\varepsilon_g^2}{2}\right)
-
W_{-1}\left(-\frac{\varepsilon_g^2}{2}\right)
\right].}{4\pi^2}
\end{equation}

The flat landscape fraction for this cost function is therefore
\begin{equation}
p_{\text{barren}} = 1 - \frac{ \frac{1}{2}
\left[
W_0\left(-\frac{\varepsilon_g^2}{2}\right)
-
W_{-1}\left(-\frac{\varepsilon_g^2}{2}\right)
\right].}{4\pi}.
\end{equation}

If each restart samples a point uniformly and independently from the parameter space, the number of jumps required to land in the trainable region is the expected number of approximate jumps required to escape the flat landscape
\begin{equation}
E[K] =\frac{1}{p_{\text{train}}}
= \frac{4\pi}{ \frac{1}{2}
\left[
W_0\left(-\frac{\varepsilon_g^2}{2}\right)
-
W_{-1}\left(-\frac{\varepsilon_g^2}{2}\right)
\right]}.
\end{equation}

This result shows that the number of required restarts scales inversely with the measure of the trainable region. This example exhibits exponentially suppressed gradients away from the trainable basin and therefore serves as an analytically tractable model of a barren-like flat optimisation landscape.

\subsection{Elliptic Bowl Landscape and Trainable Region}

To generalise the circular Gaussian landscape, we consider an anisotropic cost function of the form
\begin{equation}
C(x,y) = 1 - \exp\left(-(a x^2 + b y^2)\right),
\end{equation}
where $a$ and $b$ are positive constants. They control the curvature along the $x$ and $y$ directions, respectively. When $a \neq b$, the landscape becomes anisotropic and resembles an elliptic bowl. Such a structure naturally arises in the description of squeezed states. For this cost function, the elliptical radial coordinate $\rho \geq 0$ is defined by
\begin{equation}
    \rho^2=ax^2+by^2
\end{equation}

The gradient components are
\begin{equation}
\frac{\partial C}{\partial x} = 2a x e^{-(a x^2 + b y^2)},
\end{equation}
\begin{equation}
\frac{\partial C}{\partial y} = 2b y e^{-(a x^2 + b y^2)}.
\end{equation}

Thus, the gradient magnitude is
\begin{equation}
\|\nabla C\| = 2 e^{-\rho^2} \sqrt{a^2 x^2 + b^2 y^2}.
\label{costsqueezed}
\end{equation}

The gradient is suppressed both near and far from the origin due to the competing effects of the polynomial prefactor and exponential decay.

The trainable region is the set of parameters where the gradient magnitude exceeds a threshold $\epsilon_g$:
\begin{equation}
T_{\epsilon_g}
=
\left\{
(x,y)\in\Theta :
\|\nabla C(x,y)\|\geq\epsilon_g
\right\}.
\label{eq:exact_trainable_set}
\end{equation}

Then the gradient can be written as
\begin{equation}
\|\nabla C\| = 2 e^{-\rho^2} \sqrt{a^2 x^2 + b^2 y^2}.
\end{equation}

Using the bounds
\begin{equation}
a \rho^2 \le a^2 x^2 + b^2 y^2 \le b \rho^2 \quad \text{(assuming } a < b\text{)},
\end{equation}
we obtain
\begin{equation}
2\sqrt{a}\, \rho e^{-\rho^2}
\;\le\;
\|\nabla C\|
\;\le\;
2\sqrt{b}\,  \rho e^{-\rho^2}.
\label{inequality}
\end{equation}
Now, define 
\begin{align}
    g_a(\rho)= 2\sqrt{a}\,  \rho e^{-\rho^2}\\
    g_b(\rho)= 2\sqrt{b}\,  \rho e^{-\rho^2}
\end{align}

For $\kappa\in\{a,b\}$, define the corresponding threshold annulus, already restricted to the same domain,
\begin{equation}
\mathcal{A}_{\kappa}
=
\left\{
(x,y)\in\Theta :
2\sqrt{\kappa}\,\rho e^{-\rho^2}
\geq\epsilon_g
\right\}.
\label{eq:threshold_annulus}
\end{equation}

Equation \ref{inequality} then immediately yields the \textit{exact set inclusions}
\begin{equation}
\mathcal{A}_a
\subseteq
T_{\epsilon_g}
\subseteq
\mathcal{A}_b.
\label{eq:annular_set_inclusions}
\end{equation}

Indeed, a point satisfying the lower-bound threshold
$g_a(\rho)\geq\epsilon_g$ necessarily has a sufficiently
large actual gradient; conversely, any point with actual
gradient above the threshold must also satisfy
$g_b(\rho)\geq\epsilon_g$. These implications are rigorous
and do not require identifying the exact gradient contours
with ellipses.
For either $\kappa\in\{a,b\}$, the equation defining the boundary of
$\mathcal{A}_{\kappa}$ is
\begin{equation}
2\sqrt{\kappa}\,\rho e^{-\rho^2}=\epsilon_g.
\label{eq:annulus_boundary}
\end{equation}

When $0<\epsilon_g<\sqrt{2a/e}$, both bounding annuli are nonempty,
and their two elliptical radii are
\begin{equation}
\rho_{\kappa,-}
=
\sqrt{
-\frac{1}{2}
W_0\left(-\frac{\epsilon_g^2}{2\kappa}\right)
},
\qquad
\rho_{\kappa,+}
=
\sqrt{
-\frac{1}{2}
W_{-1}\left(-\frac{\epsilon_g^2}{2\kappa}\right)
}.
\label{eq:elliptical_radii}
\end{equation}

Here $W_0$ and $W_{-1}$ are the two real branches of Lambert's
$W$ function. Accordingly,
\begin{equation}
\mathcal{A}_{\kappa}
=
\left\{
(x,y)\in\Theta :
\rho_{\kappa,-}\leq\rho\leq\rho_{\kappa,+}
\right\}.
\label{eq:annulus_definition}
\end{equation}

The two sets are genuine elliptical annuli, whereas the exact set
$T_{\epsilon_g}$ generally has direction-dependent boundaries.
If the entire annuli lie inside $\Theta$, their areas follow from the
change of variables $u=\sqrt{a}\,x$,
$v=\sqrt{b}\,y$:
\begin{equation}
|\mathcal{A}_{\kappa}|
=
\frac{\pi}{\sqrt{ab}}
\left(
\rho_{\kappa,+}^{2}
-
\rho_{\kappa,-}^{2}
\right).
\label{eq:annulus_area}
\end{equation}

If they extend beyond the boundary, the relevant quantities are the areas of their intersections with $\Theta$. In both cases, under
independent uniform sampling on $\Theta$, the set inclusions give
rigorous probability bounds:
\begin{equation}
\frac{|\mathcal{A}_a|}{|\Theta|}
\leq
p_{\mathrm{train}}
=
\frac{|T_{\epsilon_g}|}{|\Theta|}
\leq
\frac{|\mathcal{A}_b|}{|\Theta|}.
\label{eq:trainability_probability_bounds}
\end{equation}

Here, the areas denote sets within $\Theta$ as defined in equation \ref{eq:threshold_annulus}. Writing $p_a = |\mathcal{A}_a|/|\Theta|$ and
$p_b = |\mathcal{A}_b|/|\Theta|$, Eq.~(18) implies, whenever
$p_a > 0$,
\begin{equation}
\frac{1}{p_b}
\leq
\mathbb{E}[K]
=
\frac{1}{p_{\mathrm{train}}}
\leq
\frac{1}{p_a}.
\label{eq:expected_draws_bounds}
\end{equation}

This gives a fully rigorous interpretation of the elliptical
example in terms of bounds on the expected number of
independent uniform draws. It also makes it possible to select
a conservative fixed draw budget using the lower bound $p_a$:
\begin{align}
N_{\mathrm{safe}}
=
\left\lceil
\frac{\log(1-P_{\mathrm{target}})}
{\log(1-p_a)}
\right\rceil,\\
\qquad
\Pr(\text{success by }N_{\mathrm{safe}})
\geq P_{\mathrm{target}}.
\label{eq:safe_draw_budget}
\end{align}

The gradient bounds in Equation \ref{inequality} do not, in general, identify the exact trainable region, because the gradient magnitude depends on direction at a fixed elliptical radius. Instead, they define an inner and an outer elliptical annulus, $\mathcal{A}_a$ and $\mathcal{A}_b$, such that $\mathcal{A}_a \subseteq T_{\epsilon_g} \subseteq \mathcal{A}_b$. The corresponding Lambert-$W$ radii provide explicit bounds on the area of the trainable region and, under independent uniform sampling, on the trainability probability and expected number of draws. For a bounded parameter domain, each annulus is understood to be intersected with that domain before its area is normalised.

\subsection{Gaussian Patch-Based Landscape}

Consider the cost function
\begin{equation}
C(\boldsymbol{\theta}) =
\text{plateau} -
\frac{1}{N_{\rm islands}}
\sum_{k=1}^{N_{\rm islands}}
\exp\left[- \frac{\|\boldsymbol{\theta} - \boldsymbol{\mu}_k\|^2}{2\sigma^2} \right].
\end{equation}

This landscape consists of a constant baseline with superimposed Gaussian patches centred at $\boldsymbol{\mu}_k$, each representing a trainable island. The parameter $\sigma$ controls the width of each island.

The gradient is given by
\begin{equation}
\nabla C(\boldsymbol{\theta}) =
\frac{1}{N_{\rm islands}}
\sum_{k=1}^{N_{\rm islands}}
\left[
\frac{\boldsymbol{\theta} - \boldsymbol{\mu}_k}{\sigma^2}
\exp\left(- \frac{\|\boldsymbol{\theta} - \boldsymbol{\mu}_k\|^2}{2\sigma^2} \right)
\right].
\end{equation}

Each term contributes a radially inward force toward the corresponding centre $\boldsymbol{\mu}_k$, weighted by an exponentially decaying factor.


Using gradient descent with learning rate $\eta$, the update rule is
\begin{equation}
\boldsymbol{\theta}_{t+1}
=
\boldsymbol{\theta}_t
-
\frac{\eta}{N_{\rm islands}}
\sum_{k=1}^{N_{\rm islands}}
\frac{\boldsymbol{\theta}_t - \boldsymbol{\mu}_k}{\sigma^2}
\exp\left(- \frac{\|\boldsymbol{\theta}_t - \boldsymbol{\mu}_k\|^2}{2\sigma^2} \right).
\label{grad}
\end{equation}

In the continuous gradient flow limit, the equation \ref{grad} takes the following expression:
\begin{equation}
\dot{\boldsymbol{\theta}} =
-\frac{1}{N_{\rm islands}}
\sum_{k=1}^{N_{\rm islands}}
\frac{\boldsymbol{\theta} - \boldsymbol{\mu}_k}{\sigma^2}
\exp\left(- \frac{\|\boldsymbol{\theta} - \boldsymbol{\mu}_k\|^2}{2\sigma^2} \right).
\end{equation}
Each term can be expressed as follows
\begin{equation}
 y_k=\frac{r}{N_{\rm islands}\sigma^2}
\exp(- \frac{ r^2}{2\sigma^2})
\end{equation}
where,
\begin{equation}
   r= \boldsymbol{\theta} - \boldsymbol{\mu}_k
\end{equation}

 Rearranging, it can be expressed as 

 \begin{equation}
   r_k^2 =-\sigma^2 W(-N^2_{island}\sigma^2 y_k^2)    
 \end{equation}

Here, W is the Lambert W function. If $\boldsymbol{\theta}$ is far from all $\boldsymbol{\mu}_k$, then
\begin{equation}
\exp\left(- \frac{\|\boldsymbol{\theta} - \boldsymbol{\mu}_k\|^2}{2\sigma^2} \right) \approx 0,\\
\nabla C(\boldsymbol{\theta}) \approx 0
\end{equation}
 Again when $y_l\rightarrow 0$, then
\begin{align}
\nabla C(\boldsymbol{\theta}) \approx 0,
\end{align}
Thus, each island has two regions where its gradient becomes negligible. Again, using the gradient, the trainable region is defined as
\begin{equation}
T_{\epsilon_g}
=
\left\{
\boldsymbol{\theta} :
|\nabla C| \ge \epsilon_g
\right\}.
\end{equation}

Thus, the outer and inner radii become 

\begin{align}
    r_{k,in} =\sqrt{-\sigma^2 W_{0}(-N^2_{island}\sigma^2 \epsilon^2)} \\
    r_{k,out} =\sqrt{-\sigma^2 W_{-1}(-N^2_{island}\sigma^2 \epsilon^2)}
\end{align}

Thus, we can compute the trainable region in n dimensions. The trainable region of the annulus is given as

\begin{equation}
V_{k,\mathrm{train}}
=
\frac{\pi^{n/2}}{\Gamma\left(\frac{n}{2}+1\right)}
\left(
r_{k,\mathrm{out}}^n-r_{k,\mathrm{in}}^n
\right).
\end{equation}
Assuming all the island trainable regions are inside the bounded region and all islands are identical, each island
contributes the same trainable volume. We therefore define

\begin{equation}
V_{\mathrm{island}}
\equiv V_{k,\mathrm{train}},
\qquad
\forall k.
\end{equation}

Consequently, if the trainable regions of different islands do not
overlap, the total trainable volume of the landscape is
\begin{equation}
V_{\text{train}}
=
N_{\rm islands}
\left(
V_{island}
\right).
\end{equation}

Thus, from this, we can estimate the expected number of jumps and samples for each sampling technique. Now, Assuming uniform sampling over $\boldsymbol{\theta} \in [0,2\pi]^n$, the total volume is
\begin{equation}
V_{\text{total}} = (2\pi)^n.
\end{equation}

Thus,
\begin{equation}
p_{\text{train}}
=
\frac{V_{\text{train}}}{V_{\text{total}}}
=
N_{\rm islands}
\left(
\frac{V_{island}}{(2\pi)^n}
\right).
\end{equation}

The flat landscape fraction is
\begin{equation}
p_{\text{barren}} = 1 - p_{\text{train}}.
\end{equation}

 Thus, using equation \ref{expectedjump}, the expected number of trials required to reach the trainable region is
\begin{equation}
E[K] = \frac{1}{p_{\text{train}}}
=
\frac{1}{N_{\rm islands}}
\left(
\frac{(2\pi)^n}{V_{island}}
\right).
\end{equation}

This landscape exhibits a structured flat landscape: gradient-rich regions are localised around Gaussian patches, while the remainder of the parameter space is flat. As the dimension increases, the trainable region occupies an exponentially small measure. Thus, highlighting that the proposed method is effective when the landscape contains sparse 
non-negligible trainable regions embedded within broader flat regions. If the landscape is uniformly flat under the optimisation criterion, random restarts alone cannot restore trainability. The derivation provided here is exact for a single landscape and approximate for sufficiently separated multiple islands.

\section{Experimental Design }

\subsection{Optimization Dynamics}
To understand the optimisation dynamics of the protocol, we have built two synthetic test beds and evaluated their performance under fixed sampling, a plateau window, and random restarts. In the experiments, we treat $K_{max}$ and $S$ as hyperparameters, since, for Gaussian and Cauchy Sampling, the relevant escape probability depends on the sampling distribution and the local search region. For the unscrambled Sobol sequence, the independent sampling formula does not apply, as it is a deterministic sampling technique. 

\subsubsection{Gaussian Patch Landscape\label{gaussianpatch}}

To evaluate the behaviour of the proposed SFG framework, we consider a synthetic Gaussian patch-based landscape in a two-dimensional parameter space, $(\theta \in \mathbb{R}^2)$. The landscape consists of localised trainable regions (Gaussian patches) embedded within a flat plateau.

The cost function for the optimisation process is defined as
\begin{equation}
C(\theta) = \frac{1}{M} \sum_{i=1}^{M} \left[1 - \exp\left(-\frac{|\theta - \mu_i|^2}{2\sigma^2}\right)\right],
\end{equation}

Here, M determines the number of Gaussian landscapes in the function. For convenience, in the experiment we choose $(M = 1)$, with standard deviation $(\sigma = 0.08)$, and mean $(\mu_i \sim \mathcal{U}([0, 2\pi]^2))$. The optimiser is initialised to start in a flat region. Gradient descent is performed with learning rate $(\eta = 0.01)$ for $(T = 300)$ iterations. Flat region detection is triggered over a window of size $(L = 5)$, the conditions $(|\nabla C(\theta)| < \epsilon = 0.05)$, $(\mathrm{std}(C(\theta)) < 10^{-4})$, and $(\mathbb{E}[C(\theta)] > 0.99)$ are satisfied. Upon detecting a plateau, a restart-based search is initiated. The number of candidates sampled, S, is treated as a hyperparameter and kept constant across runs. The number of sampled candidates is $S=22$. For the Gaussian and Cauchy distributions, the number of sampled points is 22. However, for Sobol sampling, the number of sampled points must be a power of 2. Since 22 is not a power of 2, we choose S, which is the closest power of 2. The search uses $(r_0 = 0.5)$, $(\delta = 0.2)$, $(r_{\max} = 3.0)$, $(K_{\max} = 25)$, and $(\epsilon_{\text{ball}} = 0.1)$.
A candidate $(\theta^)$ is accepted if $(|\nabla C(\theta^)| > \epsilon)$ and it lies outside previously explored neighbourhoods of the archive. Since the domain is mostly flat with a single region of gradient, bounding the domain and dropping sample points outside it is redundant, as the landscape would cause SFG to reject them anyway. Thus, these steps were not included in the following optimisation.

\begin{figure}
    \centering
    
    \begin{subfigure}[b]{0.3\linewidth}
        \centering
        \includegraphics[width=\linewidth]{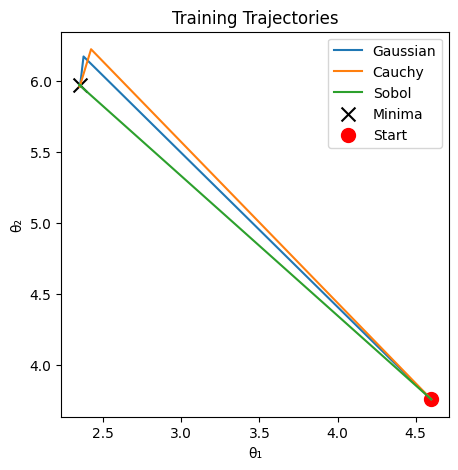}
        \caption{Training trajectory}
        \label{subfig1}
    \end{subfigure}
    \hfill
    \begin{subfigure}[b]{0.3\linewidth}
        \centering
        \includegraphics[width=\linewidth]{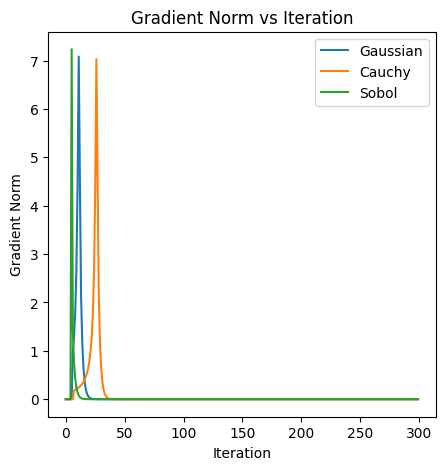}
        \caption{Cost derivative vs iteration}
        \label{subfig2}
    \end{subfigure}
    \hfill
    \begin{subfigure}[b]{0.3\linewidth}
        \centering
        \includegraphics[width=\linewidth]{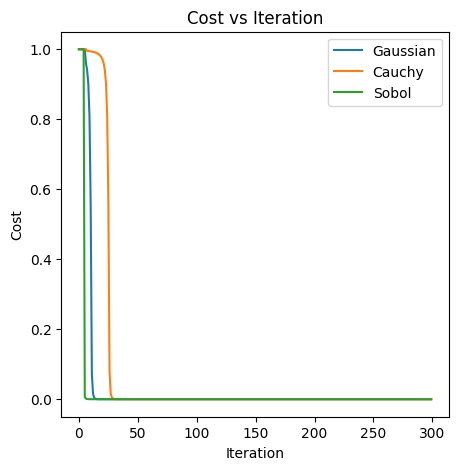}
        \caption{Cost vs iteration}
        \label{subfig3}
    \end{subfigure}

    \caption{Optimisation dynamics on the Gaussian-patch landscape. Panel (a) shows the parameter trajectory, panel (b) shows the gradient norm versus iteration, and panel (c) shows the cost function versus iteration.}
    \label{fig:gaussian_patches}
\end{figure}

The figure \ref{subfig1} shows the optimiser's trajectory for three different sampling methods. Initially, the gradient descent optimiser was trapped in a flat region, as shown in Figure \ref{subfig2}, which prompted the search for a gradient-based method. Each method successfully found a region with a gradient and escaped the flat region. Once the optimiser found the gradient, it successfully returned the minima of the cost function, as shown in Figure \ref{subfig3}.

\subsubsection{Annulus Landscape}

Similarly, we investigate the proposed SFG framework in a synthetic ring-shaped landscape in a two-dimensional parameter space, $(\theta \in \mathbb{R}^2)$. The landscape consists of a continuous annular trainable region embedded within a flat plateau.

The cost function is defined as
\begin{equation}
C(\theta) = 1 - \exp\left(-\frac{(|\theta| - r_0)^2}{2\sigma^2}\right),
\end{equation}
with radius $(r_0 = 2.0)$ and variance $(\sigma = 0.08)$. The optimiser is initialised in a flat region far from the annulus.

Gradient descent is performed with learning rate $(\eta = 0.01)$ for $(T = 300)$ iterations. Flat region detection is triggered when, over a window of size $(L = 5)$, the conditions $(|\nabla C(\theta)| < \epsilon = 0.05)$, $(\mathrm{std}(C(\theta)) < 10^{-4})$, and $(\mathbb{E}[C(\theta)] > 0.99)$ are satisfied.

Upon detecting a plateau, a restart-based search is initiated. Again, the number of samples is treated as a hyperparameter, with S=64 samples. Candidate points are generated using Gaussian, Cauchy, and Sobol sampling. The search uses $(r_0 = 0.5)$, $(\delta = 0.3)$, $(r_{\max} = 5.0)$, $(K_{\max} = 25)$, and $(\epsilon_{\text{ball}} = 0.1)$. Since $64$ is a power of 2, the number of samples generated by Sobol sampling will be $64$.

\begin{figure}
    \centering
    
    \begin{subfigure}[b]{0.3\linewidth}
        \centering
        \includegraphics[width=\linewidth]{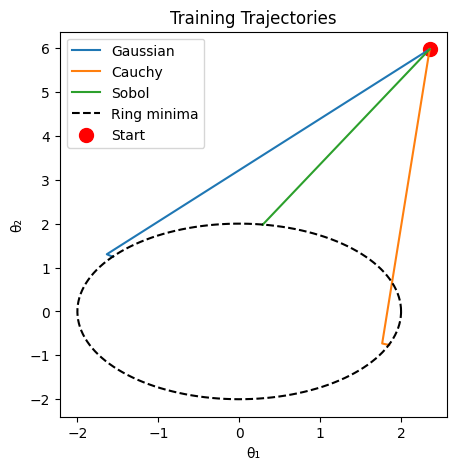}
        \caption{Training trajectory}
        \label{subfig1ring1}
    \end{subfigure}
    \hfill
    \begin{subfigure}[b]{0.3\linewidth}
        \centering
        \includegraphics[width=\linewidth]{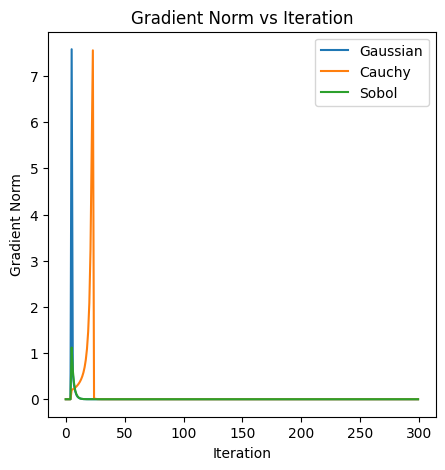}
        \caption{Cost derivative vs iteration}
        \label{subfig2ring2}
    \end{subfigure}
    \hfill
    \begin{subfigure}[b]{0.3\linewidth}
        \centering
        \includegraphics[width=\linewidth]{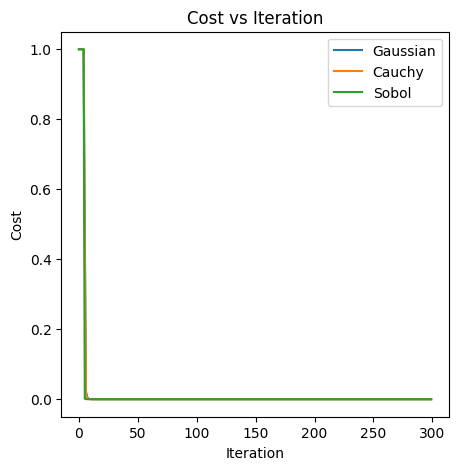 }
        \caption{Cost vs iteration}
        \label{subfig3ring3}
    \end{subfigure}

    \caption{Optimisation dynamics on the Ring Annulus: (a) shows the parameter trajectory, panel (b) shows the gradient norm versus iteration, and panel (c) shows the cost function versus iteration.}
    \label{fig:annulus}
\end{figure}

Similarly, a candidate $(\theta^)$ is accepted if $(|\nabla C(\theta^)| > \epsilon)$ and it lies outside previously explored neighborhoods. The figure \ref{subfig1ring1} shows the optimiser's trajectory for three different sampling methods. During the initial iterations, the gradient descent optimiser was trapped in a flat region, as shown in Figure \ref{subfig2ring2}, prompting the search for a gradient-based method. Each sampling method successfully identified a region with a gradient and escaped the barren plateau. Once the optimiser found the gradient, it successfully returned the minima of the cost function as shown in Figure \ref{subfig3ring3}. Similarly, since the domain is mostly flat with a single region of gradient, bounding the domain and dropping sample points that are outside it is redundant, as the landscape would cause SFG to reject them on its own. Thus, these steps were not included in the following optimisation.

\subsection{Ablation test}
\subsubsection{Gaussian patch}
We perform an ablation study to systematically evaluate the impact of key components of the proposed Search-for-Gradient framework. In particular, we analyse the role of (i) the sampling strategy, (ii) the number of candidate samples per restart ($S$), and (iii) the plateau detection window ($W$). All experiments are conducted on the Gaussian patch-based synthetic landscape, which exhibits structured flat landscapes with isolated trainable regions.

\paragraph{Sampling Strategy.}
We compare three sampling strategies: Gaussian, Cauchy, and Sobol. The test was conducted with the same hyperparameters as in Section~\ref{tab:sampling_ablation}. To perform the test, the initial point of the optimiser was held constant across all sampling strategies. The three sampling techniques achieve a 100\% success rate across the 10 independent trials and converge to the global minimum, with a final cost of $0.00 \pm 0.00$. We observe that Gaussian sampling achieves the fastest escape, requiring an average of $7.10 \pm 1.578$ jumps. The Cauchy sampling has the second-largest mean but a large standard deviation, with a mean escape time of $8.80 \pm 5.154$ jumps. The Sobol sampling has the slowest escape time, requiring $11.00 \pm 0.00$ jumps on average, with no fluctuations, as it is a deterministic sampling method. Each method successfully reached the minima of the cost function in every independent trial.

\paragraph{Effect of Candidate Count $S$.}
We vary the number of candidate samples, $S$, while keeping all other hyperparameters identical to those used in Section~\ref{gaussianpatch}. As shown in Table~\ref{tab:S_ablation_all}. Increasing $S$ does not affect the success rate. However, it consistently reduces the escape time for Gaussian and Cauchy sampling, indicating that a larger number of samples accelerates the search phase without affecting its reliability. For Gaussian sampling, the mean escape time decreases from $8.50 \pm 1.910$  for $S=64$ to $7.10 \pm 1.578$  jumps for $S=128$, and further to $5.90 \pm 0.831$  for $S=256$. Similarly, Cauchy sampling shows a reduction in escape time from $9.40 \pm 6.296$ to $8.80 \pm 5.154$, and finally to $6.30 \pm 3.9$, as $S$ increases. The Sobol sampling, on the other hand, successfully escaped the flat region on each run, but the mean escape time remained constant.

\paragraph{Effect of Plateau Detection Window $W$.}
We study the Effect of the plateau detection window, $W$, on the performance of the proposed escape mechanism. The results are presented in Table~\ref{tab:W_ablation_all}. The entire performance analysis was conducted with a sample size of $256$. As the plateau detection window $W$ varies, the success rate is unaffected across all three sampling strategies, which achieve 100\% success. The escape time has no dependence on the plateau detection window \(W\). Gaussian and Sobol sampling remain relatively stable across the tested window sizes, whereas Cauchy sampling shows substantially larger variability and a non-monotonic dependence on \(W\). These results suggest that the choice of \(W\) has a limited and sampler-dependent effect on escape dynamics for the Gaussian patch landscape.

\begin{table}
\centering
\caption{Ablation study over sampling strategies on the Gaussian patch landscape. Results are reported as mean $\pm$ standard deviation over $N = 10$ independent trials. We report success rate, escape time, and final cost.}
\label{tab:sampling_ablation}
\begin{tabular}{lccc}
\hline
Sampling Method & Success Rate & Escape Time  & Final Cost  \\
\hline
Gaussian & $1.00 \pm 0.00$ & $7.10 \pm 1.578$ & $0.00 \pm 0.00$ \\
Cauchy   & $1.00 \pm 0.00$ & $8.80 \pm 5.154$ & $0.00 \pm 0.00$ \\
Sobol    & $ 1.00 \pm 0.00$ & $11.00\pm 0.00$ & $0.00 \pm 0.00$ \\
\hline
\end{tabular}
\end{table}
\begin{table*}[t]
\centering
\caption{Effect of candidate sample size $S$ on performance across sampling strategies for the Gaussian patch landscape. Results are reported as mean $\pm$ standard deviation over $N = 10$ trials.}
\label{tab:S_ablation_all}
\begin{tabular}{lcccccc}
\hline
 & \multicolumn{2}{c}{Gaussian} 
 & \multicolumn{2}{c}{Cauchy} 
 & \multicolumn{2}{c}{Sobol} \\
$S$ 
& Success Rate & Escape Time 
& Success Rate & Escape Time 
& Success Rate & Escape Time \\
\hline

64  
& $1.00 \pm 0.00$ & $8.50 \pm 1.910$ 
& $1.00 \pm 0.00$ & $9.40 \pm 6.296$ 
& $1.00 \pm 0.00$ & $11.00 \pm 0.00$ \\

128 
& $1.00 \pm 0.00$ & $7.10 \pm 1.578$ 
& $1.00 \pm 0.00$ & $8.80 \pm 5.154$ 
& $1.00 \pm 0.00$ & $11.00 \pm 0.000$ \\

256 
& $1.00 \pm 0.00$ & $5.90 \pm 0.831$ 
& $1.00 \pm 0.00$ & $6.30 \pm 3.900$ 
& $1.00 \pm 0.00$ & $11.00 \pm 0.000$ \\

\hline
\end{tabular}
\end{table*}

\begin{table*}[t]
\centering
\caption{Effect of plateau detection window $W$ on performance across sampling strategies for the Gaussian patch landscape. Results are reported as mean $\pm$ standard deviation over $N = 10$ trials.}
\label{tab:W_ablation_all}
\begin{tabular}{lcccccc}
\hline
 & \multicolumn{2}{c}{Gaussian} 
 & \multicolumn{2}{c}{Cauchy} 
 & \multicolumn{2}{c}{Sobol} \\
$W$ 
& Success Rate & Escape Time 
& Success Rate & Escape Time 
& Success Rate & Escape Time \\
\hline

5  
& $1.00 \pm 0.00$ & $5.900 \pm 0.831$ 
& $1.00 \pm 0.00$ & $6.300 \pm 3.900$ 
& $1.00 \pm 0.00$ & $11.000 \pm 0.000$ \\

10 
& $1.00 \pm 0.00$ & $7.300 \pm 2.283$ 
& $1.00 \pm 0.00$ & $13.900 \pm 7.021$ 
& $1.00 \pm 0.00$ & $11.000 \pm 0.000$ \\

20 
& $1.00 \pm 0.00$ & $5.900 \pm 1.221$ 
& $1.00 \pm 0.00$ & $11.100 \pm 6.534$ 
& $1.00 \pm 0.00$ & $11.000 \pm 0.000$ \\

\hline
\end{tabular}
\end{table*}

\subsubsection{Annulus landscape}

We perform an ablation study on the ring-shaped landscape, where the trainable region forms a continuous manifold. The results exhibit qualitatively different behaviour from the Gaussian patch landscape.

\paragraph{Sampling Strategy}
As shown in Table~\ref{table7}, all three methods successfully escaped the flat region in every trial. Among them, Cauchy sampling exhibited the fastest escape time with a mean of $2.30$ and a standard deviation of $1.187$. Gaussian sampling has a mean escape time of $5.80$  with a standard deviation of 1.470. Sobol escape time was highest, with a mean of $14$.

\paragraph{Candidate Count}
As shown in Table~\ref{table8}, increasing the number of candidate samples, $S$, affects the escape time differently for each sampling method, while all methods maintain a 100\%  success rate. For Gaussian sampling, the mean escape time remains close to 5.5 regardless of the value of $S$. In contrast, Cauchy sampling benefits significantly from increasing $S$, with the mean escape time decreasing from $2.30\pm 1.187$ for $S=64$ to $1.7 \pm 0.640 $  for $S=128$, and further to $1.40 \pm 0.490$ for $S=256$. For Sobol sampling, increasing $S$ from 64 to 256 does not affect the escape time, and the variance remains zero as expected.

\paragraph{Effect of Plateau Detection Window W.}
As shown in Table~\ref{table9}, varying the plateau detection window, $W$, has little Effect on the performance of any of the sampling strategies. Similarly, the test was performed for a sample size of $256$. All three methods consistently achieve a 100\% success rate for all values of $W$. The escape time shows no systematic dependence on  \(W\). Gaussian and Sobol sampling remain relatively stable across the tested window sizes. These results suggest that the choice of \(W\) does not affect the dynamics in the annulus landscape.

\begin{table}
\centering
\caption{Ablation study over sampling strategies on the ring landscape. Results are reported as mean $\pm$ standard deviation over $N = 10$ independent trials. We report success rate, escape time (in iterations), and final cost.}
\label{tab:ring_ablation}
\begin{tabular}{lccc}
\hline
Sampling Method & Success Rate & Escape Time  & Final Cost \\
\hline
Gaussian & $1.00 \pm 0.00$ & $5.80  \pm  1.470$ & $0.00 \pm 0.00$ \\
Cauchy   & $1.00 \pm 0.00$ & $2.30  \pm 1.187$ & $0.00 \pm 0.00$ \\
Sobol    & $ 1.00 \pm 0.00$ & $14.00\pm 0.00$ & $0.00 \pm 0.00$ \\
\hline
\end{tabular}
\label{table7}
\end{table}

\begin{table*}[t]
\centering
\caption{Effect of candidate sample size $S$ on performance across sampling strategies for the ring landscape. Results are reported as mean $\pm$ standard deviation over $N=10$ trials.}
\label{tab:ring_S_ablation}
\begin{tabular}{lcccccc}
\hline
 & \multicolumn{2}{c}{Gaussian} & \multicolumn{2}{c}{Cauchy} & \multicolumn{2}{c}{Sobol} \\
$S$ & Success Rate & Escape Time & Success Rate & Escape Time & Success Rate & Escape Time \\
\hline
64
& $1.00 \pm 0.00$ & $5.80 \pm 1.470$
& $1.00 \pm 0.00$ & $2.30 \pm 1.187$
& $1.00 \pm 0.00$ & $14.00 \pm 0.00$ \\

128
& $1.00 \pm 0.00$ & $6.40 \pm 1.020$
& $1.00 \pm 0.00$ & $1.70 \pm 0.640$
& $1.00 \pm 0.00$ & $13.00 \pm 0.00$ \\

256
& $1.00 \pm 0.00$ & $5.10 \pm 0.943$
& $1.00 \pm 0.00$ & $1.40 \pm 0.490$
& $1.00 \pm 0.00$ & $13.00 \pm 0.00$ \\
\hline
\end{tabular}
\label{table8}
\end{table*}

\begin{table*}[t]
\centering
\caption{Effect of plateau detection window $W$ on performance across sampling strategies for the ring landscape with a candidate sample size of $S=256$. Results are reported as mean $\pm$ standard deviation over $N=10$ trials.}
\label{tab:ring_W_ablation}

\begin{tabular}{lcccccc}
\hline
 & \multicolumn{2}{c}{Gaussian} & \multicolumn{2}{c}{Cauchy} & \multicolumn{2}{c}{Sobol} \\
$W$ & Success Rate & Escape Time & Success Rate & Escape Time & Success Rate & Escape Time \\
\hline

5
& $1.00 \pm 0.00$ & $5.10 \pm 0.943$
& $1.00 \pm 0.00$ & $1.40 \pm 0.490$
& $1.00 \pm 0.00$ & $13.00 \pm 0.00$ \\

10
& $1.00 \pm 0.00$ & $5.10 \pm 1.136$
& $1.00 \pm 0.00$ & $1.20 \pm 0.400$
& $1.00 \pm 0.00$ & $13.00 \pm 0.00$ \\

20
& $1.00 \pm 0.00$ & $5.10 \pm 0.700$
& $1.00 \pm 0.00$ & $1.50 \pm 0.500$
& $1.00 \pm 0.00$ & $13.00 \pm 0.00$ \\

\hline
\end{tabular}
\label{table9}
\end{table*}

\subsection{Comparative Study with Standard Optimisers}

To test the effectiveness of the proposed framework, we conduct a comparative study against standard optimisation methods, including Gradient Descent (GD), Adam, Simultaneous Perturbation Stochastic Approximation (SPSA), and Random Restart (RR).

\subsubsection{Comparative Study on Gaussian Patch Landscape \label{GPL_optimizer}}

We evaluate the performance of the proposed Search-for-Gradient (SFG) framework against standard optimisation methods, including Gradient Descent (GD), Adam, SPSA, and Random-Restart Gradient Descent (RR-GD), on the Gaussian patch landscape. This landscape is particularly challenging because the trainable region is highly localised and occupies only a small fraction of the parameter space. In contrast, the remaining region is nearly flat with vanishing gradients.

For a consistent comparison of optimisation progress, we adopt a common evaluation-count convention, with budgets of 900 and 1500 evaluation units in two separate trials. For GD and Adam, each gradient evaluation was counted as one evaluation. Since GD and Adam do not require evaluating any cost function, they will not contribute to the budget. On the other hand, SPSA estimates the gradient using two cost-function evaluations per iteration. Thus, each SPSA iteration consumes two cost-function evaluations from the common budget. For RR-GD, the total number of gradient evaluations across all restarts was constrained to a fixed budget of 100 iterations. Moreover, each iteration requires evaluating a single gradient. Similarly, for SFG, both the gradient evaluations performed during the gradient descent phase and the candidate gradient evaluations carried out during the search phase contributed to the total evaluation budget. For SFG, the sample size of 64 was considered for each iteration. So each rounds in the search phase required evaluating 64 gradient points. It is important to emphasise that this convention provides a common accounting framework for comparing the optimisation methods; it does not imply that a full gradient evaluation and a cost-function evaluation have equivalent computational costs.

The hyperparameters of the optimisers were held constant across all experiments. Gradient Descent used a learning rate of 0.01. Adam had a learning rate of 0.01, with first- and second-moment decay coefficients of 0.9 and 0.999, respectively, and a numerical stability parameter of $10^{-8}$. SPSA, on the other hand, was implemented using a fixed learning rate of 0.05 and a perturbation magnitude of 1.0, where the perturbation vector was sampled independently from a Bernoulli distribution over $\{-1,+1\}$ at every iteration. Random Restart Gradient Descent used the same learning rate, 0.01, as GD, with each restart allowed up to 100 gradient steps while maintaining the common evaluation budget.

The proposed SFG framework initially performs gradient descent optimisation with a learning rate of 0.01 until the average gradient norm over the previous five iterations falls below 0.05, indicating that the optimiser has entered a flat region. The search phase is initiated when these conditions are satisfied, with an initial search radius of 0.5. For each search radius, 64 candidate points are sampled from a Gaussian distribution centred at the current iterate, and the candidate with the largest gradient norm is selected. If no candidate exceeds the prescribed gradient threshold, the search radius is increased by 0.3. A maximum of 20 radius expansions is permitted before terminating the search phase at a maximum radius of 3.

Table~\ref{tab:gaussian_results} summarises the final optimisation results. GD, Adam, SPSA, and RR-GD all fail to reach the global minimum, converging instead to a final cost of $1.00 \pm 0.00$. In contrast, the proposed SFG framework consistently converges to the global minimum, achieving a final cost of $0.00 \pm 0.00$ in every trial.

The optimisation dynamics are illustrated in Figures~\ref{gaussiancost}, \ref{gaussiangradientnorm}, and \ref{gaussiantrajectory}. As shown in Figure~\ref{gaussiancost}, the cost remains essentially unchanged for GD, Adam, SPSA, and RR-GD throughout the optimisation, indicating that these methods are unable to make meaningful progress toward the optimum. Correspondingly, Figure~\ref{gaussiangradientnorm} shows that the gradient norm rapidly approaches zero and remains negligible for the entire optimisation. Since these optimisers rely solely on local gradient information, the absence of a usable gradient prevents any effective parameter updates. Random Restart GD also fails because each restart is overwhelmingly likely to initialise in another flat region, where the optimiser again encounters vanishing gradients and terminates without approaching the localised basin of attraction.

In contrast, the SFG framework detects prolonged periods of negligible gradients using knowledge of the mean and variance, and explicitly enters a search phase. Rather than relying on local gradient information at the current point, it samples candidate points in the surrounding parameter space using Gaussian sampling. It selects the location with the largest gradient magnitude. This mechanism enables the optimiser to escape the flat landscape and discover informative descent directions. As illustrated in Figure~\ref{gaussiantrajectory}, SFG successfully traverses the plateau and reaches the localised Gaussian basin, after which conventional gradient descent rapidly converges to the global minimum. This transition is also reflected in Figure~\ref{gaussiangradientnorm}, where a sharp increase in the gradient norm occurs immediately after the search phase, followed by a rapid decay as the optimiser approaches the optimum. Table~\ref {tab:budget_table} shows that the outcome does not change for both the 900 and 1500 budgets.

These results show that the proposed method overcomes one of the limitations of conventional methods on landscapes with predominantly flat regions. It does so by entering the search phase, in which it seeks informative gradients rather than relying solely on local ones. SFG can locate and exploit regions with non-zero gradients, enabling successful optimisation where existing gradient-based methods fail.

\begin{table}
\centering
\begin{tabular}{lcc}
\hline
\textbf{Optimizer} & \textbf{Final Cost} & \textbf{Outcome} \\
\hline
GD      & $1.00 \pm 0.00$ & Failed \\
Adam    & $1.00 \pm 0.00$ & Failed \\
SPSA    & $1.00 \pm 0.00$ & Failed \\
RR-GD   & $1.00 \pm 0.00$ & Failed \\
SFG     & $0.00 \pm 0.00$ & Success \\
\hline
\end{tabular}
\caption{Final cost comparison on the Gaussian patch landscape for GD, ADAM, SPSA, RR-GD and SFG optimisers for a budget of 1500 iterations.}
\label{tab:gaussian_results}
\end{table}

\subsubsection{Comparative Study on Annulus Landscape}
We evaluate the performance of the proposed SFG framework on the annulus landscape and compare it with GD, Adam, SPSA, and RR-GD. Unlike the Gaussian patch landscape, the global minima of the annulus landscape form a continuous ring, making the optimisation problem fundamentally different. For the experiment, the hyperparameters are kept the same as those in section \ref{GPL_optimizer} with a sample size of 64 in each round. We adopt the same accounting framework as Gaussian patch landscape for annulus landscape.

Table~\ref{tab:ring_results} , Table \ref{tab:budget_table} summarizes the final optimization results. GD, Adam, and SPSA fail to converge to the global minimum, each reaching a final cost of $1.00 \pm 0.00$. In contrast, both RR-GD and the proposed SFG framework converge to the global minimum, yielding a final cost of $0.00 \pm 0.00$ with a budget of 1500. However, when the evaluation budget is reduced to 900, RR-GD fails to converge and terminates with a final cost of $1.00 \pm 0.00$. In contrast, SFG continues to converge successfully to the global minimum. 

\begin{figure}
    \centering

    \begin{subfigure}{0.45\linewidth}
        \centering
        \includegraphics[width=\linewidth]{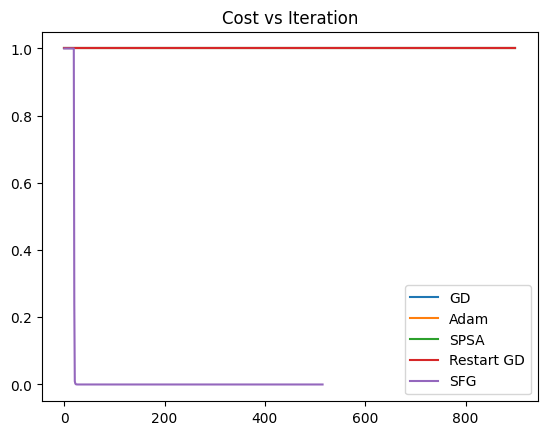}
        \caption{Gradient landscape Cost vs iterations}
        \label{gaussiancost}
    \end{subfigure}
    \hfill
    \begin{subfigure}{0.45\linewidth}
        \centering
        \includegraphics[width=\linewidth]{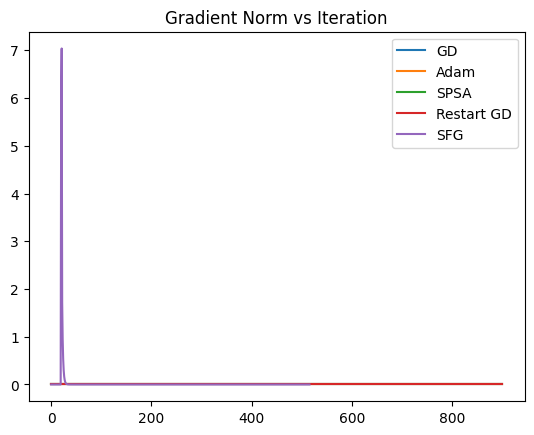}
        \caption{Gradient landscape gradient norm vs iterations}
        \label{gaussiangradientnorm}
    \end{subfigure}
    \hfill
    \begin{subfigure}{0.45\linewidth}
        \centering
        \includegraphics[width=\linewidth]{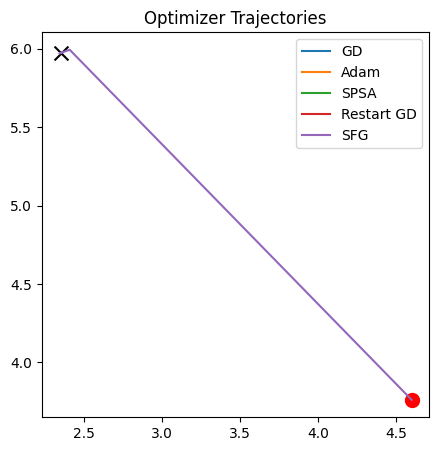}
        \caption{Trajectory of every optimiser in Gaussian Landscape}
        \label{gaussiantrajectory}
    \end{subfigure}

    \vspace{0.5cm}

    \begin{subfigure}{0.45\linewidth}
        \centering
        \includegraphics[width=\linewidth]{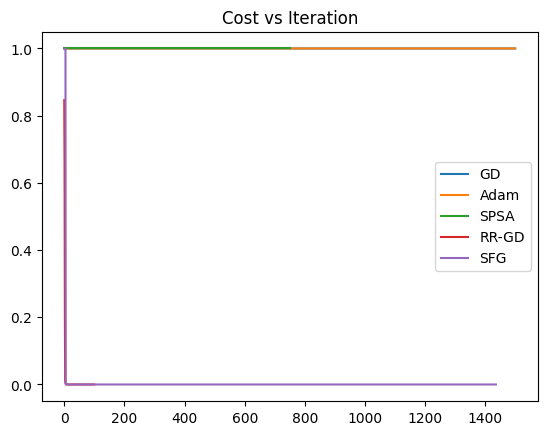}
        \caption{Annulus landscape cost vs iterations}
        \label{annuluscost}
    \end{subfigure}
    \hfill
    \begin{subfigure}{0.45\linewidth}
        \centering
        \includegraphics[width=\linewidth]{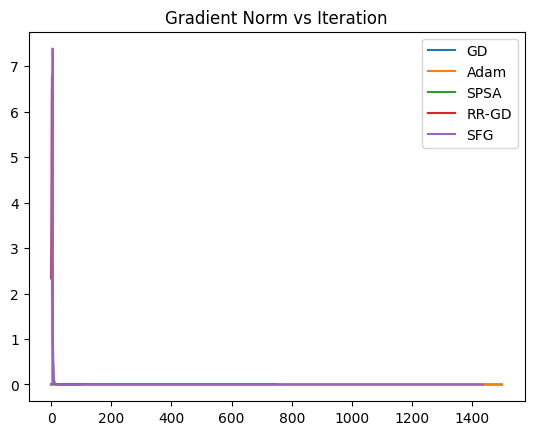}
        \caption{Annulus landscape gradient norm vs iterations }
        \label{annulusGradientcost}
    \end{subfigure}
    \hfill
    \begin{subfigure}{0.45\linewidth}
        \centering
        \includegraphics[width=\linewidth]{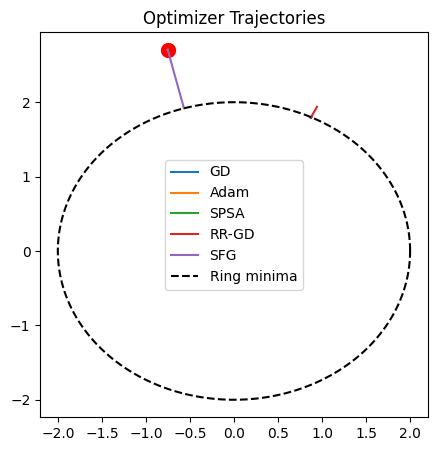}
        \caption{Trajectory of every optimiser on annulus landscape}
        \label{trajectoryofannulus}
    \end{subfigure}

    \caption{Comparison of the proposed Search-for-Gradient (SFG) framework with Gradient Descent (GD), Adam, SPSA, and Random Restart Gradient Descent (RR-GD) on the Gaussian patch and annulus landscapes. (a)--(c) show the cost evolution, gradient norm, and optimisation trajectories on the Gaussian patch landscape, respectively. (d)--(f) present the corresponding results for the annulus landscape. On the Gaussian patch landscape, only SFG successfully escapes the flat region and converges to the global minimum. In contrast, GD, Adam, SPSA, and RR-GD remain trapped in regions of vanishing gradients. On the annulus landscape, SFG and RR-GD successfully converge to the ring of global minima under a sufficiently large evaluation budget. At the same time, GD, Adam, and SPSA fail due to the absence of informative gradient signals.}
    \label{fig:four_subfigures}
\end{figure}

The optimisation dynamics are illustrated in Figures~\ref{annuluscost}, \ref{annulusGradientcost}, and \ref{trajectoryofannulus}. As shown in Figure~\ref{annuluscost}, the cost remains nearly constant for GD, Adam, and SPSA throughout the optimisation, indicating that these methods become trapped in the flat region surrounding the ring. This behaviour is further confirmed by Figure~\ref{annulusGradientcost}, where the gradient norm rapidly decreases to nearly zero and remains negligible for the remainder of the optimisation. Without informative gradients, these first-order optimisation methods cannot make meaningful progress toward the global minimum.

Random Restart Gradient Descent successfully converges under the larger evaluation budget because repeated random initialisations increase the probability that at least one starting point lies sufficiently close to the ring of minima. Once initialised within the basin of attraction of the ring, conventional gradient descent rapidly converges to the optimum. The corresponding Trajectory is shown in Figure~\ref{trajectoryofannulus}. However, when the evaluation budget is reduced, the number of allowable restarts decreases, thereby reducing the probability of initialising near the ring and consequently leading to optimisation failure.

The proposed SFG framework also converges successfully to the ring minimum. After detecting that the optimiser has entered a flat region, SFG enters the local search to identify a point with a stronger signal for optimisation, then resumes gradient descent. As shown in Figures~\ref{annulusGradientcost} and \ref{trajectoryofannulus}, this search phase enables the optimiser to escape the plateau and converge reliably to the nearest point on the ring.

Comparing the annulus results with the plots obtained on the Gaussian patch landscape (Figures~\ref{gaussiancost}--\ref{gaussiantrajectory}). This shows the influence of the landscape geometry. While RR-GD succeeds on the annulus landscape because the global minima form a continuous manifold with a relatively large basin of attraction, it fails on the Gaussian patch landscape, where the minimum is confined to a highly localised region. In contrast, the proposed SFG framework successfully converges on both landscapes, demonstrating its robustness across optimisation problems with different geometries.

\begin{table}[h]
\centering
\begin{tabular}{lcc}
\hline
\textbf{Optimizer} & \textbf{Final Cost} & \textbf{Outcome} \\
\hline
GD      & $1.00 \pm 0.00 $& Failed \\
Adam    & $1.00 \pm 0.00 $& Failed \\
SPSA    & $1.00 \pm 0.00$ & Failed \\
RR-GD   & $0.00\pm 0.00 $& Success \\
SFG     & $0.00 \pm 0.00$ & Success \\
\hline
\end{tabular}
\caption{Final cost comparison on the ring (annulus) landscape for GD, ADAM, SPSA, RR-GD and SFG optimisers for a budget of 1500 iterations.}
\label{tab:ring_results}
\end{table}

\begin{table}[t]
\centering
\caption{Comparison of optimisation methods under different evaluation budgets on the Gaussian patch and ring landscapes.}
\label{tab:budget_comparison}
\begin{tabular}{lccc}
\hline
\textbf{Optimizer} & \textbf{Budget} & \textbf{Gaussian Patch} & \textbf{Ring} \\
\hline
GD      & 900  & Failed  & Failed \\
GD      & 1500 & Failed  & Failed \\
\hline
Adam    & 900  & Failed  & Failed \\
Adam    & 1500 & Failed  & Failed \\
\hline
SPSA     & 900  & Failed  & Failed \\
SPSA     & 1500 & Failed  & Failed \\
\hline
RR-GD   & 900  & Failed  & Failed \\
RR-GD   & 1500 & Failed  & Success \\
\hline
SFG     & 900  & Success & Success \\
SFG     & 1500 & Success & Success \\
\hline
\end{tabular}
\label{tab:budget_table}
\end{table}

\subsection{Real Quantum Circuit}
\begin{figure}
    \centering
    
    \begin{subfigure}[b]{0.48\linewidth}
        \centering
        \includegraphics[width=\linewidth]{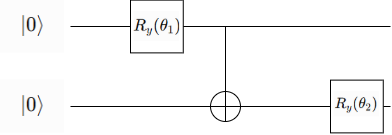}
        \caption{Quantum circuit for SFG.}
        \label{fig:quantumsfg_circuit}
    \end{subfigure}
    \hfill
    \begin{subfigure}[b]{0.48\linewidth}
        \centering
        \includegraphics[width=\linewidth]{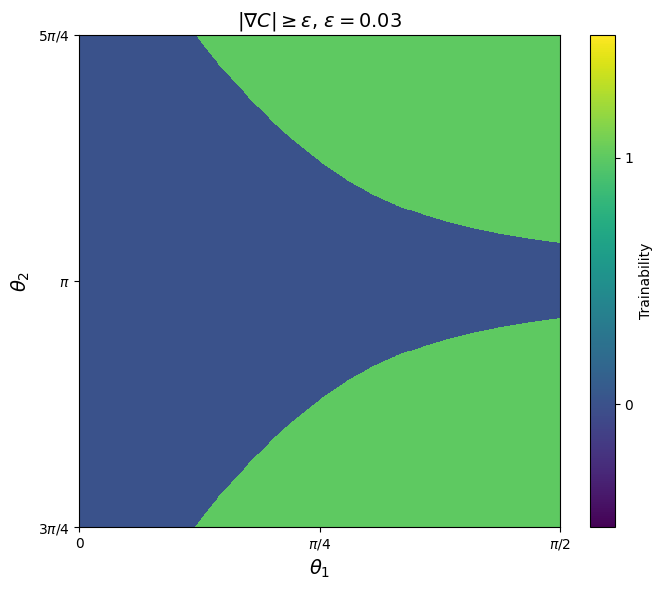}
        \caption{Optimisation landscape.}
        \label{fig:realcircuitlandscape}
    \end{subfigure}
    
    \caption{Quantum circuit implementation of the SFG protocol and its corresponding optimisation landscape.}
    \label{fig:quantumcircuitsfg_results}
\end{figure}

 To test SFG in the real quantum circuit, we performed an optimisation on a cost function derived from a real quantum circuit, as shown in subfigure \ref{fig:quantumsfg_circuit}. The domain of the cost function is $(\theta_1,\theta_2)\in\left[0,\frac{\pi}{2}\right]\times\left[\frac{3\pi}{4},\frac{5\pi}{4}\right]$. The truncation of the domain helps us treat the landscape as a structured flat landscape, making it suitable to test SFG. The domain has a $56\%$ flat region with a threshold value of $0.03$ as shown in subfigure \ref{fig:realcircuitlandscape}. The cost function for the circuit is defined as follows:
 \begin{equation}
C = 1 - \left|\bra{11}\ket{\psi}\right|^2
= 1 - \sin^2\left(\frac{\theta_1}{2}\right)
    \cos^2\left(\frac{\theta_2}{2}\right).
\end{equation}
In this function, SFG is used to escape the flat region and restart the optimisation. The SFG optimiser is implemented with initial points at \(\boldsymbol{\theta}_0=(0.07,\pi)\). The gradient descent uses a learning rate of \(\eta=0.01\) for up to \(T=300\) iterations.

Similarly, for this circuit, a plateau is detected using a window of \(L=5\) consecutive iterations. Three conditions are simultaneously required: the gradient norm must satisfy \(\|\nabla C(\boldsymbol{\theta})\|<\epsilon\), with \(\epsilon=0.03\); the mean cost over the detection window must exceed \(\zeta=0.95\); and the standard deviation of the cost over the same window must be below \(\sigma_{\mathrm{cost}}=10^{-4}\). Once these conditions are satisfied, SFG initiates its restart search.

The restart search begins with an initial radius of \(r_0=0.05\). If an attempted search fails to identify an acceptable candidate, the radius is increased by \(\delta=0.05\) with an upper bound on the radius $r_{max}=3.5$. Similarly, the total number of samples is treated as a hyperparameter. A total of \(128\) candidate data points are generated at each search attempt. A candidate is accepted when its gradient norm exceeds the plateau threshold, \(\|\nabla C(\boldsymbol{\theta}^{*})\|>\epsilon\), and it lies outside previously explored restart neighbourhoods defined by a ball of radius \(\epsilon_{\mathrm{ball}}=0.1\). During sampling, points outside the domain are discarded.

\begin{figure}
    \centering

    \begin{subfigure}[b]{0.48\linewidth}
        \centering
        \includegraphics[width=\linewidth]{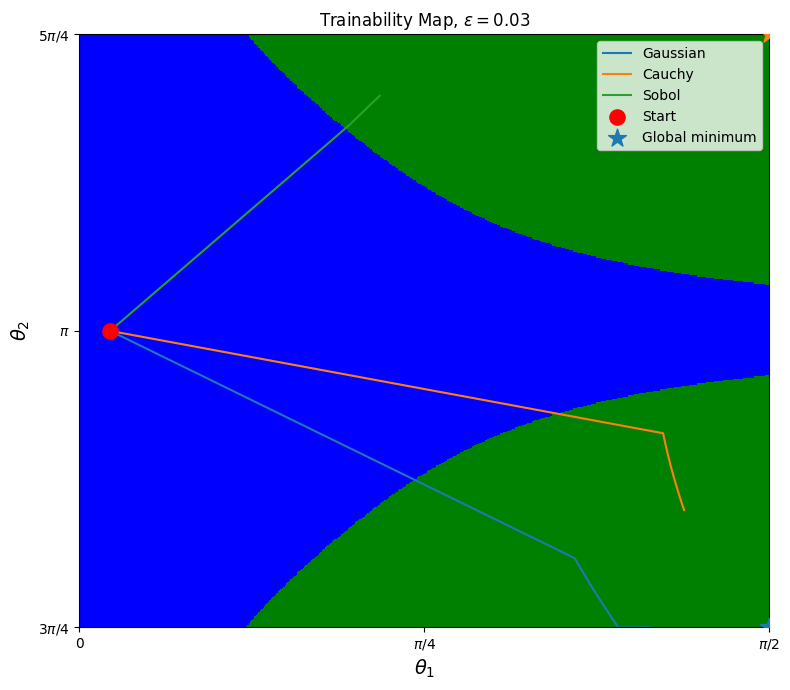}
        \caption{Optimisation trajectories of the different sampling methods.}
        \label{fig:realcircuittrajectory}
    \end{subfigure}
    \hfill
    \begin{subfigure}[b]{0.48\linewidth}
        \centering
        \includegraphics[width=\linewidth]{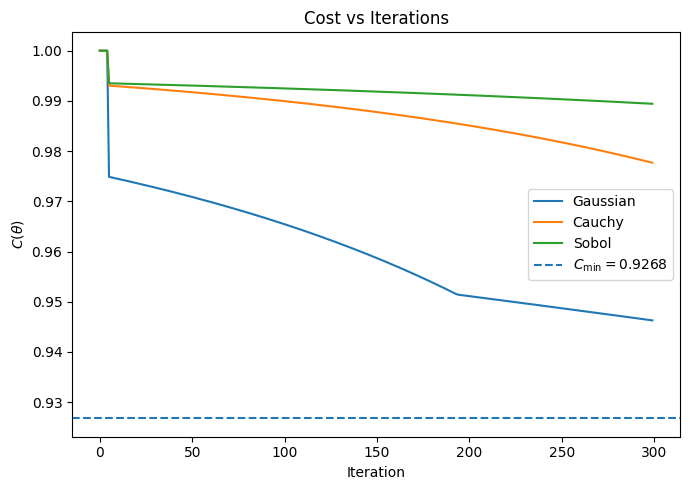}
        \caption{Cost function versus optimisation iteration.}
        \label{fig:realcircuitcostvsiter}
    \end{subfigure}

    \vspace{2mm}

    \begin{subfigure}[b]{0.48\linewidth}
        \centering
        \includegraphics[width=\linewidth]{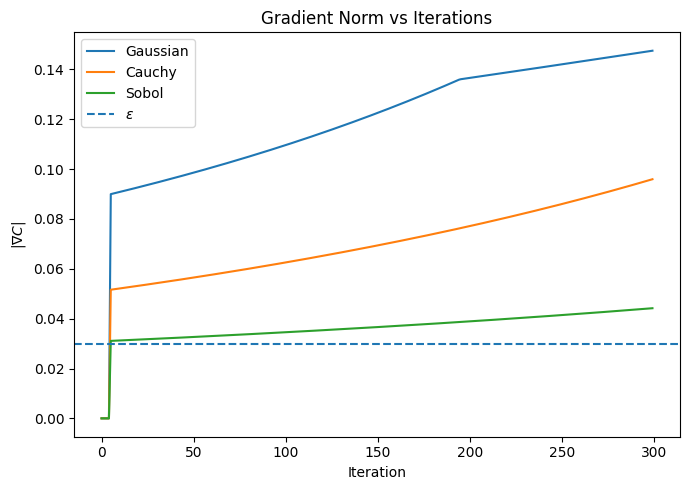}
        \caption{Gradient norm versus optimisation iteration.}
        \label{fig:realcircuitgradvsiter}
    \end{subfigure}

    \caption{
        Performance of SFG for the optimisation of the cost landscape
        derived from the quantum circuit. 
        (a) Optimisation trajectories generated using different sampling
        strategies. 
        (b) Evolution of the cost function with optimisation iterations.
        (c) Evolution of the gradient norm, illustrating the convergence
        behaviour of the optimiser.
    }
    \label{fig:realcircuitresults}
\end{figure}

The $K_{max}$ parameters are chosen using a target sampling probability \(P_{\mathrm{target}}=0.98\) and an estimated total sampling probability \(p_{\mathrm{total}}=0.56\). The probability threshold for radius adaptation is set to \(\delta_p=0.10\). The maximum number of restart-search attempts is set to twice the quantity computed in equation \ref{k_max}, yielding $K_{max}=14$, which is implemented directly in the experimental configuration. This choice of $K_{max}$ is a conservative implementation choice, rather than a direct consequence of the probabilistic formula. Since the landscape is a truncated version of the whole landscape, sampling methods may yield data points outside the domain. In those cases, we drop those points that are outside the domain. For each optimiser, the escape was successful, as shown in subfigure \ref{fig:realcircuittrajectory}. The figure shows that each optimiser successfully escaped the flat region. The subfigures \ref{fig:realcircuitcostvsiter} and \ref{fig:realcircuitgradvsiter} show that the gradient has recovered after initially becoming negligible. Also, the cost value is gradually approaching the minimum. The plots illustrate that the optimiser was initially stuck in a flat region. After SFG, the optimiser successfully escaped the barren region and restarted its gradient descent.

\section{Discussion and conclusion}

The results demonstrate that the limitation of gradient-based optimisation in variational quantum algorithms is that local update rules cannot discover regions where meaningful gradient information exists. In such settings, optimisation becomes fundamentally a search problem rather than a purely local descent process.

The proposed framework addresses this limitation. This framework is a mechanism that, when a flat region is encountered, shifts to a search phase in which it actively searches for parameter regions with non-vanishing gradients. Instead of attempting to improve optimisation within a flat landscape, the method seeks to relocate the optimiser to a region where optimisation is feasible.

The main contribution of this work is the notion of a structured flat landscape. Unlike fully random landscapes, where gradients vanish uniformly across the parameter space, many practical variational circuits exhibit heterogeneous structure, with isolated or extended regions of non-negligible gradient embedded within predominantly flat regions. The experimental results on Gaussian patches and annulus landscapes support this view. In both cases, the optimisation dynamics are dominated by rare transitions from flat regions to trainable regions, after which standard gradient descent becomes effective. In addition, such a landscape was observed in the real circuit, and SFG was tested in that landscape, successfully escaping the flat region and restarting the optimisation.

The experimental results show that the SFG is governed primarily by the geometry of the trainable regions rather than by the sampling strategy alone. Different sampling methods excel in different landscape geometries, which suggests that geometry-aware sampling is a promising direction for future optimisation. In the Gaussian landscape, Gaussian sampling achieved the fastest and most consistent escape. In contrast, in the annulus landscape, Cauchy sampling provided the quickest escape due to its ability to generate larger exploratory jumps. Sobol sampling successfully escaped the flat landscape in all experiments but generally exhibited higher escape time. Also, since it is unscrambled Sobol sampling, the escape time had zero variance. These results indicate that the choice of sampling strategy should be matched to the geometry of the optimisation landscape rather than assuming a universally superior method.

Despite its advantages, the proposed framework has several limitations. First, its performance relies on the existence of non-negligible trainable regions. In landscapes that are uniformly barren under the chosen cost function, no search-based strategy can recover useful gradients. Second, the method incurs additional computational overhead due to sampling and evaluating candidate points.

Finally, the framework allows it to integrate with existing variational algorithms and extend them in several directions. In particular, designing circuits with a structured landscape could further enhance trainability. These methods provide ways to perform optimisation guided by both gradients and search over the parameter space.

\section{Future Work}

The \textit{Search-for-Gradient} framework has several promising directions for future research. One such direction is to design an ansatz that resembles a structurally flat landscape. Extending the framework to noisy intermediate-scale quantum (NISQ) devices by testing under noise models and hardware constraints could enhance robustness in practical settings. Another important direction is estimating the total plateau probability, $p_{\mathrm{total-ball}}$. An important limitation of the current framework is that $p_{\mathrm{total-ball}}$ is treated as a hyperparameter, which could lead to oversampling or undersampling within the search neighbourhood. Developing an adaptive method to estimate $p_{total-ball}$  during optimisation, without introducing significant computational overhead, could both improve the efficiency of the framework and provide a deeper theoretical understanding of the geometry of flat optimisation landscapes. Applying the method to real quantum machine learning tasks—such as classification, generative modelling, or kernel-based learning—would help validate its utility beyond synthetic landscapes. Additionally, the sampling strategy could be made adaptive to dynamically select between Gaussian, Cauchy, or Sobol sampling based on feedback from the cost landscape. Finally, the framework is adaptable to classical machine learning problems with structured, flat loss landscapes, such as deep networks with saturated activations or overparameterised models.

\appendix

\section{Theorem: Structured flat Landscape $\sigma$-algebra}
\begin{theorem}[Structured flat Landscape $\sigma$-algebra]
Let $\Theta$ be a bounded parameter space with a disjoint decomposition
\begin{equation}
\Theta = \left(\bigcup_{i \in I} B_i \right) \cup \left(\bigcup_{k \in K} T_k \right),
\end{equation}
where each $B_i$ denotes a barren region, and each $T_k$ denotes a trainable region. 
Assume that all sets are pairwise disjoint:
\begin{align}
B_i \cap B_j &= \varnothing \quad (i \neq j), \\
T_k \cap T_l &= \varnothing \quad (k \neq l), \\
B_i \cap T_k &= \varnothing \quad \forall i \in I,\; k \in K.
\end{align}

Define the collection
\begin{equation}
\mathcal{A}
=
\left\{
\bigcup_{i \in I'} B_i \;\cup\;
\bigcup_{k \in K'} T_k
\;:\;
I' \subseteq I,\;
K' \subseteq K
\right\}.
\end{equation}
Then $\mathcal{A}$ is a $\sigma$-algebra on $\Theta$.
\label{sigma}
\end{theorem}

\begin{proof}
We verify the three properties of a $\sigma$-algebra.

\textbf{(i) Empty set:}  
Choosing $I' = \varnothing$ and $K' = \varnothing$ gives
\[
\emptyset \in \mathcal{A}.
\]

\textbf{(ii) Closed under complement:}  
Let
\[
A = \bigcup_{i \in I'} B_i \cup \bigcup_{k \in K'} T_k \in \mathcal{A}.
\]
Since the sets $\{B_i\}$ and $\{T_k\}$ form a disjoint decomposition of $\Theta$,
\[
A^c
= \Theta \setminus A
= \bigcup_{i \in I \setminus I'} B_i
\cup
\bigcup_{k \in K \setminus K'} T_k,
\]
which is again of the same form, hence $A^c \in \mathcal{A}$.

\textbf{(iii) Closed under countable unions:}  
Let $\{A_n\}_{n=1}^\infty$ be a sequence of sets in $\mathcal{A}$ where
\[
A_n = \bigcup_{i \in I_n} B_i \cup \bigcup_{k \in K_n} T_k.
\]
Then
\[
\bigcup_{n=1}^{\infty} A_n
=
\bigcup_{i \in \left(\cup_n I_n\right)} B_i
\cup
\bigcup_{k \in \left(\cup_n K_n\right)} T_k,
\]
which is again in $\mathcal{A}$.

Therefore, $\mathcal{A}$ satisfies all properties of a $\sigma$-algebra on $\Theta$.
\end{proof}

\ifCLASSOPTIONcaptionsoff
\newpage
\fi

\bibliographystyle{IEEEtran}

\bibliography{IEEE}

\vfill

\end{document}